\documentclass[journal]{IEEEtran}

\ifCLASSINFOpdf
\else
\fi

\usepackage{cite}
\usepackage{graphicx}
\usepackage{caption}
\usepackage{physics, amsmath,amssymb,amsfonts,algpseudocode}
\usepackage{upgreek}
\usepackage{optidef}
\usepackage{multirow}
\usepackage{array}
\usepackage{stfloats}
\usepackage{amsthm}
\usepackage{lipsum}
\usepackage{mathtools}
\usepackage{cuted}
\usepackage[linesnumbered, ruled]{algorithm2e}
\usepackage{blindtext}
\usepackage{amsfonts}
\usepackage{gensymb}

\usepackage{textcomp}
\usepackage{todonotes}
\begin{document}

\title{Active STAR-RIS Empowered Edge System for Enhanced Energy Efficiency and Task Management}

\author{Pyae~Sone~Aung,
        Kitae~Kim,~\IEEEmembership{Student~Member,~IEEE,}
        Yan~Kyaw~Tun,~\IEEEmembership{Member,~IEEE,}
        Zhu~Han,~\IEEEmembership{Fellow,~IEEE,}        and~Choong~Seon~Hong,~\IEEEmembership{Fellow,~IEEE,}
\thanks{Pyae Sone Aung, Kitae Kim and Choong Seon Hong are with the Department of Computer Science and Engineering, Kyung Hee University, Yongin-si, Gyeonggi-do 17104, Republic of Korea (e-mail: pyaesoneaung@khu.ac.kr; glideslope@khu.ac.kr; cshong@khu.ac.kr).}
\thanks{Yan Kyaw Tun is with Department of Electronic Systems, Aalborg University, A . C. Meyers Vænge 15, 2450 København (e-mail: ykt@es.aau.dk).}
\thanks{Zhu Han is with the Electrical and Computer Engineering Department, University of Houston, Houston, TX 77004, and the Department of Computer Science and Engineering, Kyung Hee University, Yongin-si, Gyeonggi-do 17104, Republic of Korea (email: hanzhu22@gmail.com).}}

\maketitle

\begin{abstract}
The proliferation of data-intensive and low-latency applications has driven the development of multi-access edge computing (MEC) as a viable solution to meet the increasing demands for high-performance computing and storage capabilities at the network edge. Despite the benefits of MEC, challenges such as obstructions cause non-line-of-sight (NLoS) communication to persist. Reconfigurable intelligent surfaces (RISs) and the more advanced simultaneously transmitting and reflecting (STAR)-RISs have emerged to address these challenges; however, practical limitations and multiplicative fading effects hinder their efficacy. We propose an active STAR-RIS-assisted MEC system to overcome these obstacles, leveraging the advantages of active STAR-RIS. The main contributions consist of formulating an optimization problem to minimize energy consumption with task queue stability by jointly optimizing the partial task offloading, amplitude, phase shift coefficients, amplification coefficients, transmit power of the base station (BS), and admitted tasks. Furthermore, we decompose the non-convex problem into manageable sub-problems, employing sequential fractional programming for transmit power control, convex optimization technique for task offloading, and Lyapunov optimization with double deep Q-network (DDQN) for joint amplitude, phase shift, amplification, and task admission. Extensive performance evaluations demonstrate the superiority of the proposed system over benchmark schemes, highlighting its potential for enhancing MEC system performance. Numerical results indicate that our proposed system outperforms the conventional STAR-RIS-assisted by 18.64\% and the conventional RIS-assisted system by 30.43\%, respectively.
\end{abstract}


\IEEEpeerreviewmaketitle

\section{Introduction}

\subsection{Background and Motivations}
The surge in the adoption of data-intensive and low-latency applications, including video streaming services, social media platforms, and online gaming, has catalyzed the development of multi-access edge computing (MEC) as a plausible resolution. MEC is a distributed computing paradigm that brings computational resources and storage capabilities closer to end-users and their devices at the edge of the network \cite{taleb2017multi}. In MEC architecture, computing resources are deployed in proximity to where data is generated or consumed, typically at the base station (BS) or access point of a cellular network. This enables low-latency and high-bandwidth processing of data, as well as efficient delivery of services. Nevertheless, several obstacles, such as tall structures, may hinder the achievement of line-of-sight (LoS) communication between the BS and end-users, which leads to performance degradation.

On the other hand, reconfigurable intelligent surface (RIS) has emerged as a promising technology, gaining significant attention from academia and industry. RIS constitutes a planar meta-surface comprising inexpensive passive elements, programmable via integrated electronic circuits to adjust the incoming electromagnetic field toward a desired direction \cite{huang2019reconfigurable}. Although RIS addresses challenges related to establishing LoS links, its deployment is constrained by limitations in coverage range and practicality, necessitating both transmitter and receiver to be situated on the same side. To surmount this limitation, research focus has shifted towards simultaneously transmitting and reflecting (STAR)-RIS. As implied by its name, STAR-RIS has the capability to reflect the incident signal on the same side while simultaneously transmitting it on the opposite side \cite{mu2021simultaneously}. Consequently, STAR-RIS offers better degree of freedom (DoF) and capacity gains for manipulating signal propagation, thereby enabling to 360 \degree coverage. However, in practical scenarios, the direct link is not completely blocked by obstacles, resulting in the multiplicative fading effect. This effect entails the cumulative path loss along the cascade from the transmitter to the STAR-RIS to the receiver. Unlike additive fading, multiplicative fading arises from the product of the path losses between the transmitter and the STAR-RIS, and between the STAR-RIS and the receiver. Therefore, achieving significant capacity gains with passive STAR-RIS configurations becomes unattainable. In response to this constraint, a new architecture for STAR-RIS termed active STAR-RIS, has been introduced. Unlike its passive counterpart, active STAR-RIS can mitigate the substantial path loss encountered in both the reflected and transmitted links through its amplification capabilities \cite{zhang2022active}.

\subsection{Research Contributions}
Taking into account the previously mentioned challenges and motivations, this paper aims to propose an active STAR-RIS-assisted MEC system, leveraging the benefits offered by both active STAR-RIS and task queue stability. The main contributions of this paper can be summarized as follows.
\begin{itemize}    
    \item We propose an active STAR-RIS-assisted MEC system and formulate the optimization problem to minimize the energy consumption of user devices, active STAR-RIS, and BS by jointly optimizing the partial task offloading, amplitude, phase shift coefficients, amplification coefficients, transmit power of the BS and admitted tasks.
    \item Given the non-convex nature of the formulated problem and the challenges caused by the dynamic and unpredictable nature of the environment and task arrivals, we propose deep reinforcement learning (DRL) as a solution strategy. Nevertheless, implementing DRL for the entire problem is infeasible due to the excessive number of action spaces. Therefore, we initially decompose the main problem into three sub-problems. 
    \item Afterwards, we propose sequential fractional programming to address the transmit power control problem, standard convex optimization technique to address the partial task offloading problem, and Lyapunov optimization with double deep q-network (DDQN) to address the joint amplitude, phase shift, amplification, and task admission problem, respectively.
    \item Finally, a comprehensive performance evaluation is executed to illustrate the effectiveness of our proposed system compared to benchmark schemes, such as active STAR-RIS with full offloading, passive STAR-RIS, active-RIS and passive-RIS, respectively.
\end{itemize}

The remaining portion of the study is organized as follows: we provide an overview of the related literature in Section \ref{relatedworks}. Section \ref{systemmodel} represents the system model and problem formulation. Furthermore, Section \ref{solution} examines the solution strategy, while Section \ref{evaluation} provides the performance evaluations. Finally, Section \ref{conclusion} concludes our paper.

\section{Related Works}\label{relatedworks}

\subsection{RIS-Assisted Communications}
A comprehensive overview of of the literature associated with RIS-assisted communications is discussed in this sub-section \cite{huang2019reconfigurable, zhang2020reconfigurable, yang2021energy, aung2023energy, chen2020reconfigurable}. In \cite{huang2019reconfigurable}, the authors proposed to maximize the energy efficiency in the RIS-assisted wireless communication by jointly optimizing the phase shifts and downlink transmit power, while complying to the the constraints of maximum power and minimal quality of service. In \cite{zhang2020reconfigurable}, the authors studied the the impact of the number of phase shifts on achievable data rate in uplink RIS-assisted communication system. In \cite{yang2021energy}, the authors investigated the energy efficiency maximization with distributed RISs-assisted downlink wireless communication system by joint optimization of RIS on-off, phase and transmit power. In \cite{aung2023energy}, the authors proposed the energy-efficient downlink communications networks via multiple aerial RISs by jointly optimizing the aerial RISs deployment, reflecting elements on-off states, phase shifts and power control. In \cite{chen2020reconfigurable}, the authors studied the RIS-assisted device-to-device communication by joint optimization of transmitted power and discrete phase shift in order to maximize sum-rate.

Additionally, several researches have been conducted on RIS-assisted MEC systems \cite{he2023joint, qin2023joint, mao2022reconfigurable, zhi2022active, peng2022active}. In \cite{he2023joint}, the authors proposed a multi-server MEC system leveraged by multiple distributed RISs to maximize the task completion rate by joint optimization of user association, reflecting coefficients of RISs, receive beamforming vector and computation resource allocation. In \cite{qin2023joint}, the authors aimed to maximize the energy efficiency for the RIS-assisted unmanned aerial vehicle (UAV)-enabled MEC system under the non-orthogonal multiple access (NOMA) protocol. In \cite{mao2022reconfigurable}, the authors developed a secure computation task offloading technique for RIS-assisted MEC system which includes optimizing RIS deployment to prevent leaking information to malicious eavesdropper. Nonetheless, the works mentioned above only considered passive RIS, which limits the notable capacity gains due to the double fading effect from the cascaded transmitter-RIS-receiver links. To solve this, the authors in \cite{zhi2022active} explored the impact of amplification on active RIS and its potential enhancement on achievable rates compared to passive RIS configurations within identical power constraints. Moreover, in \cite{peng2022active}, the authors proposed methods to maximize the computational latency of the MEC system through the utilization of active RIS. However, all of the aforementioned works only considered conventional RIS-assisted systems, which limit the flexibility and coverage range for the users positioned behind the RIS.

\subsection{STAR-RIS-Assisted Communications}
A comprehensive overview of the literature associated with STAR-RIS-assisted communications is addressed in this sub-section \cite{mu2021simultaneously, wu2021coverage}. In \cite{mu2021simultaneously}, the authors introduced the practical protocols for the functionality of STAR-RIS in both unicast and multicast wireless communication scenarios. In \cite{wu2021coverage}, the authors investigated how much STAR-RIS can improve the coverage range relative to RIS in both orthogonal multiple access (OMA) and NOMA systems.

Furthermore, numerous studies have performed on STAR-RIS-assisted MEC systems \cite{zhang2023resource, liu2023star}. In \cite{zhang2023resource}, the authors proposed to maximize the energy consumption of all users in the STAR-RIS-assisted MEC system by joint optimization of reflection and transmission coefficients, time, transmit power, and offloading size. In \cite{liu2023star}, the authors aimed to maximize the computation rate for all the users with the binary reflection and transmission amplitude coefficients for the STAR-RIS-aided MEC system. Nevertheless, the aforementioned studies failed to account for the prior knowledge of computation task demands, a condition often impractical in numerous mobile computing contexts characterized by the stochastic arrival of tasks. Furthermore, conventional STAR-RIS suffers from the potential multiplicative fading effect.

\subsection{Active STAR-RIS-Assisted Communications}
A comprehensive overview of of the literature associated with active STAR-RIS-assisted communications is addressed in this sub-section \cite{xu2023active, papazafeiropoulos2024two, ma2023active}. In \cite{xu2023active}, the authors conducted a performance analysis for the active STAR-RIS-aided two users wireless communication system, comparing it to the passive STAR-RIS. The authors of \cite{papazafeiropoulos2024two} investigated the spectral efficiency maximization problem for the active STAR-RIS-assisted massive multiple-input multiple-output (mMIMO) system under several protocols. Similarly, in \cite{ma2023active}, the authors examined to maximize the spectral efficiency and improve the quality of service (QoS) of the active STAR-RIS-assisted cell-free mMIMO system system. However, the aforementioned works have not been carried out on active STAR-RIS-assisted MEC system. In such a system, active STAR-RIS can mitigate the inherent path loss in reflected and transmitted links through amplification. Consequently, it can solve the multiplicative fading effect, thereby facilitating notable capacity gains for the users within MEC systems.

\section{System Model and Problem Formulation}\label{systemmodel}
\begin{figure}[t]
	\includegraphics[width=\linewidth]{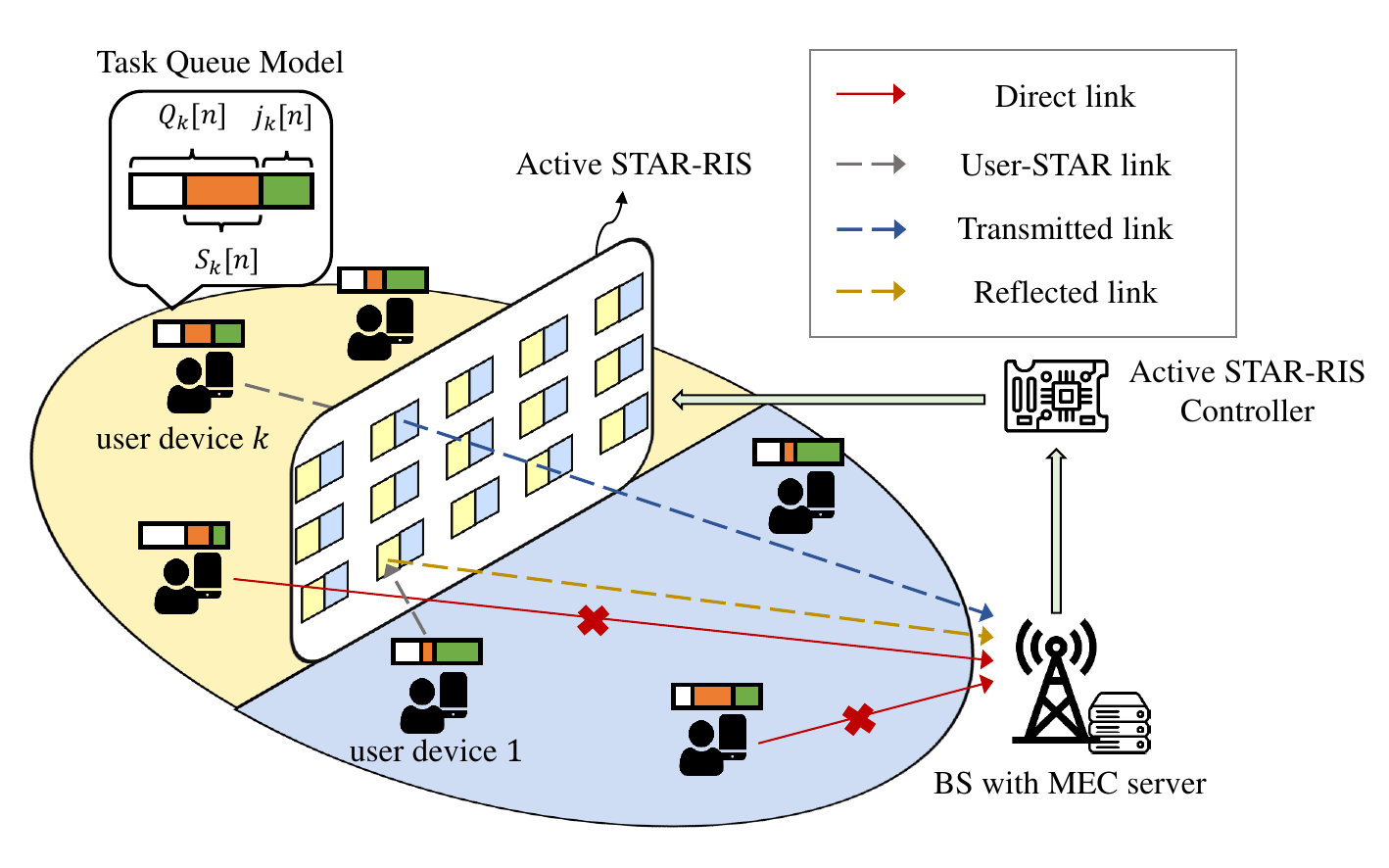}
	\caption{System model of active STAR-RIS-assisted MEC system.}
	\label{sm}
\end{figure}

Fig.~\ref{sm} depicts an active STAR-RIS-assisted MEC system which consists of a BS with multiple antennas $B$, implemented with the MEC server, a set $\mathcal{K}$ of $K$ users each equipped with single-antenna mobile device, and an active STAR-RIS. For each user device $k \in \mathcal{K}$, there is a set of tasks that needed to be computed. However, due to the restricted energy and computation capabilities, some of these tasks must be offloaded to the MEC server in order to reduce the latency and energy consumption. Considering the highly efficient computational power of the MEC server and the relatively small size of the output, it is justifiable to ignore the computational time at the MEC server and the time needed to download the result \cite{zhang2023resource}. Nonetheless, due to obstruction such as tall buildings, the lack of signal connectivity between the MEC server and user devices prevents the establishment of line-of-sight (LoS) communication links. To facilitate this, an active STAR-RIS, comprising an array of $\mathcal{M} = \{1, 2 \dots, M\}$ elements and amplifiers \cite{zhi2022active}, is deployed to enhance the communication between the MEC server and user devices. Each element either reflects or transmits the incoming signal toward the desired direction depending on the position of the receiver. Therefore, the signals of all user devices for task offloading ought to be reflected/transmitted to the BS particularly through the active STAR-RIS. We denote the user devices situated in the reflected region as $R$ and those in the transmitted region as $T$, and the total number of user devices is represented by $K=R+T$. The total service time for MEC server, denoted as $\hat{T}$ is partitioned into $n \in \mathcal{N} = \{1, 2, \dots, N\}$ time slots with equal length, represented as $\Lambda=\frac{\hat{T}}{N}$. 

\subsection{Computation Model}
Each task assigned to user device $k$ at time slot $n$ can be denoted by a tuple $\{S_k[n], C_k, D_k^{\max}[n] \}$, in which $S_k[n]$ represents the size of the input data in bits, $C_k$ represents the number of computation resources in CPU cycles required to calculate a single bit of data, and $D_k^{\max}[n]$ represents the maximum tolerable delay for completing the task. Due to their limited computational capacity and energy constraints, user devices cannot manage to execute all computation tasks locally \cite{aung2024aerial}. It is therefore necessary to divide some of the tasks to the MEC server to help execute computation. This partitioning is facilitated by the inherent bit-wise independence of the tasks, allowing certain segments of a larger tasks to be processed locally while others to efficiently offloaded to the MEC server. Therefore, we denote the variable $o_k[n]$ as the ratio of tasks to be offloaded to the MEC server. Consequently, $(1-o_k[n])$ becomes the ratio of the tasks to be computed locally. Afterward, the time required for use device $k$ to compute the tasks locally can be obtained as
\begin{equation}
    t_k^{\mathrm{loc}}[n] = \frac{(1-o_k[n]) S_k[n] C_k}{f_k[n]}, \forall{k} \in \mathcal{K},
\end{equation}
where $f_k[n]$ is the allocated computation frequency of user device $k$ at time slot $n$. The energy consumption of  user device $k$ for local computation at time slot $n$ can be calculated as
\begin{equation}
    E_k^{\mathrm{loc}}[n] = P_k^{\mathrm{loc}}[n] t_k^{\mathrm{loc}}[n], \forall{k} \in \mathcal{K},
\end{equation}
where $P_k^{\mathrm{loc}}[n] = \kappa f_k^3[n]$ is the power consumption of the CPU in user device $k$, with coefficient $\kappa$ which depends on the chip design of user device.

\subsection{Communication Model}
In this study, we assume perfect channel state information (CSI) is known at the BS. The acquisition and refinement of CSI are vital aspects in wireless communication system. However, these features are considered beyond the scope of this study, and can be attainable through established advanced channel estimation techniques from \cite{wu2021channel}, \cite{chen2023channel}, and \cite{papazafeiropoulos2024two}. Since there is no direct connection between the user devices and the BS, the communication links consists of user devices to active STAR-RIS links, and active STAR-RIS to BS (reflected/transmitted) links, respectively. As in \cite{mu2021simultaneously}, both of these channel links are modeled as Rician fading channel. Therefore, the channel gain for user device $k$ to active STAR-RIS can be expressed as
\begin{align}
&\begin{aligned}
    \mathbf{g}_{k,M}[n] = &\sqrt{\mu d_{k,M}^{-\delta}[n]} \sqrt{\frac{K}{1+K}}\mathbf{g}_{k,M}^{\mathrm{LoS}}[n] + \\ &\sqrt{\frac{1}{1+K}}\mathbf{g}_{k,M}^{\mathrm{NLOS}} \in \mathbb{C}^{1 \times |\mathcal{M}|},\\
\end{aligned}
\end{align}
where $\mu$ represents the channel gain at reference distance 1 m, $d_{k,M}[n]$ is the Euclidean distance between user $k$ and active STAR-RIS at time slot $n$, $\delta \geq 2$ represents the path loss exponent, $K$ is the Rician factor, $\mathbf{g}_{k,M}^{\mathrm{LoS}}[n]$ is the deterministic LoS component between user device $k$ and active STAR-RIS at time slot $n$, established by the azimuth angle-of-arrival (AoA) of the link, $\mathbf{g}_{k,M}^{\mathrm{NLoS}}$ is the non LoS component with a circularly-symmetric independent and identically distributed complex normal distribution, respectively. Correspondingly, the channel gain for active STAR-RIS to the BS can be expressed as
\begin{align}
&\begin{aligned}
    \mathbf{G}_{M,B}[n] = &\sqrt{\mu d_{M,B}^{-\delta}[n]} \sqrt{\frac{K}{1+K}}\mathbf{g}_{M,B}^{\mathrm{LoS}}[n] + \\
    &\sqrt{\frac{1}{1+K}}\mathbf{g}_{M,B}^{\mathrm{NLOS}} \in \mathbb{C}^{|\mathcal{M}| \times |\mathcal{B}|}.\\
\end{aligned}
\end{align}

Furthermore, we define the reflection coefficient as ${\theta}_r^m$ and transmission coefficient as ${\theta}_t^m$ of each element $m$ of active STAR-RIS, respectively, which can be obtained as
\begin{equation}
    {\theta}_r^m[n] = {\beta}_r^m[n] e^{j{\phi}_r^m[n]}, 
\end{equation}
\begin{equation}
    {\theta}_t^m[n] = {\beta}_t^m[n] e^{j{\phi}_t^m}[n],
\end{equation}
where ${\beta}_r^m[n], {\beta}_t^m[n] \in [0,1]$ and ${\phi}_r^m[n], {\phi}_t^m[n] \in [0,2\pi)$ represent the amplitudes and phase shifts of the $m$-th element of reflection and transmission coefficients of active STAR-RIS at time slot $n$, respectively. We utilize the energy-splitting model for the active STAR-RIS, assuming all elements function in both transmission and reflection modes \cite{mu2021simultaneously}. Hence, with no energy dissipation by active STAR-RIS, the energy of the incident signal on each element is split into the energies of the transmitted and reflected signals. Therefore, due to law of conservation of energy, we have ${\beta}_r^m[n] + {\beta}_t^m[n] = 1$. For simplicity, we assume discrete values and independent adjustment for phase shift coefficients of transmission and reflection, satisfying the following constraint, i.e., ${\phi}^m_{\lambda} = \frac{\iota \pi}{2^{b-1}}, \lambda \in \{r,t\},$ where $b$ is the controllable bit, and $\iota \in \{0,1,\dots,2^b-1\}$ \cite{zhao2022simultaneously}. Consequently, the reflection and transmission coefficient matrices may then be achieved as
\begin{equation}
    \mathbf{\Theta}_r[n] = \operatorname{diag}({\beta}^1_r[n] e^{j{\phi}^1_r[n]}, \dots, {\beta}^M_r[n] e^{j{\phi}^M_r[n]})
    \in \mathbb{C}^{|\mathcal{M}| \times |\mathcal{M}|},
\end{equation}
\begin{equation}
    \mathbf{\Theta}_t[n] = \operatorname{diag}({\beta}^1_t[n] e^{j{\phi}^1_t[n]}, \dots, {\beta}^M_t[n] e^{j{\phi}^M_t[n]})
    \in \mathbb{C}^{|\mathcal{M}| \times |\mathcal{M}|}.
\end{equation}

The implementation of active STAR-RIS system integrates each element with active amplifier operated by the power supply. This will result in the amplification of the received signal, followed by its simultaneous reflection/transmission towards the user devices \cite{ma2023optimization}. Therefore, with amplification coefficient, $\sqrt{\alpha^m}[n]$ of the $m$-th element at time slot $n$, the amplification matrix can be expressed as $\mathbf{A}[n] = \operatorname{diag}(\sqrt{\alpha^1}[n], \sqrt{\alpha^2}[n], \dots, \sqrt{\alpha^M}[n])$. The received signal by the BS is given by
\begin{equation}
\resizebox{\hsize}{!}{$
    \mathbf{y}[n] = \left( \sum_{k=1}^{K} \mathbf{g}_{k,M}[n] x_k[n] + \boldsymbol{\nu} \right) \mathbf{\Theta}_{\lambda}[n] \mathbf{A}[n] \mathbf{G}_{M,B}[n] + \boldsymbol{w}, \lambda \in \{r,t\},
$}
\end{equation}
where $x_k[n] = \sqrt{p_k}[n] s_k$ is the signal transmitted from user device $k$ with transmit beamforming power $p_k$ and unit information symbol $s_k$, $\boldsymbol{\nu}\sim\mathcal{CN}(\mathbf{0},\dot{\sigma}^2\mathbf{I}_M)$ is the noise produced by the active STAR-RIS, $\boldsymbol{w}\sim\mathcal{CN}(\mathbf{0},\ddot{\sigma}^2\mathbf{I}_B)$ is the noise at the BS, respectively. Thus, the uplink signal-to-interference-plus-noise-ratio (SINR) of user device $k$ can be expressed as
\begin{equation}
    \gamma_k[n] = \frac{|\mathbf{g}_k[n]|^2 p_k[n]}{\resizebox{.7\hsize}{!}{$ \sum_{i \neq k}^K |\mathbf{g}_j[n]|^2 p_j[n] + \dot{\sigma}^2|| \mathbf{\Theta}_\lambda[n] \mathbf{A}[n] \mathbf{G}_{M,B}[n] ||^2 + \ddot{\sigma}^2 $}}, \forall k \in \mathcal{K},
\end{equation}
where $\mathbf{g}_k[n] = \mathbf{g}_{k,M}[n] \mathbf{\Theta}_\lambda[n] \mathbf{A}[n] \mathbf{G}_{M,B}[n]$ represents the equivalent channel from user device $k$ to the BS.

Afterwards, at the $n$-th time slot, the achievable data rate can be achieved as
\begin{equation}
    r_k[n] = W \log_{2}(1+\gamma_k[n]) , \forall k \in \mathcal{K},
\end{equation}
where $W$ is the bandwidth. Since the user devices have restricted computation capacity, some of the computation tasks are required to be offloaded to the MEC server. Hence, the transmission delay for offloading ratio of tasks $o_k$ from user device $k$ can be obtained as
\begin{equation}
    t^{\mathrm{off}}_k[n] = \frac{o_k[n] S_k[n]}{r_k[n]}, \forall k \in \mathcal{K}.
½\end{equation}

The energy consumption of user device $k$ for offloading tasks transmission at time slot $n$ can be calculated as
\begin{equation}
    E^{\mathrm{off}}_k[n] = P^{\mathrm{off}}_k[n] t^{\mathrm{off}}_k[n] = \frac{p_k[n] o_k[n] S_k[n]}{W \log_{2}(1+\gamma_k[n])}, \forall k \in \mathcal{K}.
\end{equation}

Thereby, the total energy consumption of user device can be attained as
\begin{flalign}
    &E^{\mathrm{Tot}}_k[n] = E^{\mathrm{loc}}_k[n] + E^{\mathrm{off}}_k[n] \\
    &= (1-o_k[n]) \kappa f_k^2[n] S_k[n] C_k + \frac{p_k[n] o_k[n] S_k[n]}{W \log_{2}(1+\gamma_k[n])}, \forall k \in \mathcal{K}. \nonumber
\end{flalign}

\subsection{Task Queue Model}

During time slot $n$, the computation task arrivals of user device $k$ is denoted as $J_k[n]$ in bits, and each user device $k$ has the capacity to handle a certain subset of the incoming tasks, $j_k[n] \leq J_k[n]$ \cite{yang2022dynamic}. The task management system within each user device is structured around a queue mechanism, where the tasks are dealt with a first-in-first-out (FIFO) sequence. Consequently, the queue backlog length in bits for input data during task execution at time slot $n$ can be established as
\begin{equation}
    Q_k[n+1] = \max\{Q_k[n]-S_k[n] ,0\}+j_k[n], \forall k \in \mathcal{K}.
\end{equation}

The task queue stability constraint can be derived based on the Lyapunov optimization theory as follows \cite{neely2022stochastic}.
\begin{equation}
    \lim_{N \rightarrow \infty}\frac{1}{N} \sum_{n=1}^{N} \sum_{k=1}^{K} \mathbb{E} \{ Q_k[n] \} < \infty,
\end{equation}
where the expectation is considered in regards to system factors, which includes the channel models, arrival of computation tasks.

\subsection{Problem Formulation}
Following our system model, we aim to minimize energy consumption of user devices, active STAR-RIS, and BS in the active STAR-RIS-assisted MEC system. By jointly optimizing the partial task offloading, amplitude, phase shift coefficients, amplification coefficients, transmit power of the BS and admitted tasks, we formulate our optimization problem as follows:
\begin{mini!}|b|[1]
	{\boldsymbol{o}, \boldsymbol{\beta}_{\lambda}, \boldsymbol{\phi}_{\lambda}, \boldsymbol{\alpha}, \boldsymbol{p}, \boldsymbol{j}}  {\sum_{n=1}^{N} \sum_{k=1}^{K} E^{\mathrm{Tot}}_k (\boldsymbol{o}, \boldsymbol{\beta}_{\lambda}, \boldsymbol{\phi}_{\lambda}, \boldsymbol{\alpha}, \boldsymbol{p}, \boldsymbol{j}) [n] \label{c0}}
	{\label{OF}}{\textbf{P:}}
    \addConstraint{\sum_{k=1}^{K} ||\mathbf{A}[n] \mathbf{G}_{M,B}[n]||^2 p_k + \dot{\sigma}^2 ||\mathbf{A}[n]||_F^2 \leq P_{\mathrm{RIS}}[n]}\label{c1}
    \addConstraint{0 \leq p_k[n] \leq p_k^{\max}[n], \forall{k} \in \mathcal{K}, \forall{n} \in \mathcal{N}}\label{c2}
	\addConstraint{t_k^{\mathrm{loc}}[n] + t_k^{\mathrm{off}}[n] \leq D_k^{\max}, \forall{k} \in \mathcal{K}, \forall{m} \in \mathcal{M}}\label{c3}
    \addConstraint{\sum_{k=1}^{K} o_k[n] S_k[n] C_k \leq C^0, \forall{m} \in \mathcal{M}}\label{c4}
    \addConstraint{j_k[n] \leq J_k[n], \forall{k} \in \mathcal{K}, \forall{n} \in \mathcal{N}}\label{c5}
    \addConstraint{S_k[n] \leq Q_k[n], \forall{k} \in \mathcal{K}, \forall{n} \in \mathcal{N}}\label{c6}
    \addConstraint{\lim_{N \rightarrow \infty}\frac{1}{N} \sum_{n=1}^{N} \sum_{k=1}^{K} \mathbb{E} \{ Q_k[n] \} < \infty}\label{c7}
    \addConstraint{\beta^{r}_m[n] + \beta^{t}_m[n] = 1, \forall{m} \in \mathcal{M}, \forall{n} \in \mathcal{N}}\label{c8}
    \addConstraint{0 \leq \phi^{t}_m[n], \phi^{r}_m[n] < 2\pi, \forall{m} \in \mathcal{M}, \forall{n} \in \mathcal{N}}\label{c9}
    \addConstraint{0 \leq o_k[n] \leq 1, \forall{k} \in \mathcal{K}, \forall{n} \in \mathcal{N},}\label{c10} 
\end{mini!}
where $\boldsymbol{o}[n] = \{o_1[n], \dots, o_k[n]\}, \forall{n} \in \mathcal{N}$ is the partial task offloading vector, $\boldsymbol{\beta}_\lambda[n] = \{\beta^1_\lambda[n], \dots, \beta^M_\lambda[n]\}$, $\lambda \in \{r, t\}, \forall{n} \in \mathcal{N}$ is the amplitude vector, $\boldsymbol{\phi}_\lambda[n] = \{\phi^1_\lambda[n], \dots, \phi^M_\lambda[n]\}$, $\lambda \in \{r, t\}, \forall{n} \in \mathcal{N}$ is the phase shift coefficient vector, $\boldsymbol{\alpha}[n] = \{\sqrt{\alpha}^1[n], \dots, \sqrt{\alpha}^M[n]\}, \forall{n} \in \mathcal{N}$ is the amplification coefficient vector, $\boldsymbol{p}[n] = \{p_1[n], \dots, p_k[n]\}, \forall{n} \in \mathcal{N}$ is the transmit power vector, and $\boldsymbol{j}[n] = \{j_1[n], \dots, j_k[n]\}, \forall{n} \in \mathcal{N}$ is the admitted task vector, respectively. Furthermore, constraints (\ref{c1}) and (\ref{c2}) represent the total power budget for active STAR-RIS and each user device, respectively, constraint (\ref{c3}) represents the maximum tolerable latency for task completion, constraint (\ref{c4}) represents the computation resources available at the MEC server, constraint (\ref{c5}) represents the task admission constraint, (\ref{c6}) ensures that the executed data cannot exceed the queue backlog length, constraints (\ref{c8}) and (\ref{c9}) represent the feasible amplitude values and phase shift values for the coefficients of transmission and reflection of active STAR-RIS, and finally constraint (\ref{c10}) represents the ratio of offloaded tasks. The formulated problem is non-convex, due to the couplings between control variables in the objective function and constraints. Traditional optimization methods are incapable to fully address problem $\textbf{P1}$ in its entirety since there exists numerous local optimal solutions. Therefore, we propose DRL as a method to address the challenges posed by the dynamic and unpredictable nature of mobile device environments and task arrivals, characterized by a vast array of potential actions and uncertain outcomes. DRL is particularly advantageous because it excels in environments where obtaining an exact model is challenging, allowing it to learn optimal strategies directly from interactions with the environment. Additionally, DRL adapts continually to changes, which makes it highly effective in dynamic settings where the conditions and requirements frequently evolve.

\section{Solution Approach}\label{solution}
First of all, employing DRL across the entire optimization problem is impractical, primarily because the combined action spaces for all control variables in problem $\textbf{P1}$ would be excessively large, complicating the learning process for the DRL agent. Therefore, we adopt a three-step approach iteratively to mitigate this issue and streamline the DRL process. Considering previous studies such as \cite{lee2020deep} and \cite{hassan2024spaceris}, it is feasible to solve one or more decision variables via an optimization approach while determining the rest of the variables through a machine learning approach to minimize computational demands.\footnote{Disregarding the interdependencies and interactions between variables in an optimization issue by decoupling control variables can result in a decline in performance. However, this approach simplifies the optimization process, making it more manageable and computationally efficient, especially in complex systems.} Initially, we address the transmit power control problem using a sequential fractional programming method in Section \ref{tpcp} and convex optimization for the partial task offloading problem independently to reduce the action space in Section \ref{ptop}. Subsequently, we tackle the remaining optimization problem through the Lyapunov optimization with DRL approach in Section \ref{jointprob}.

\subsection{Transmit Power Control Problem}\label{tpcp}
With fixed $\boldsymbol{o}, \boldsymbol{\beta}_{\lambda}, \boldsymbol{\phi}_{\lambda}, \boldsymbol{\alpha}, \boldsymbol{j}$, the first sub-problem can be rewritten as

\begin{mini!}|b|[1]
	{\boldsymbol{p}} {\sum_{n=1}^{N} \sum_{k=1}^{K}E^{\mathrm{Tot}}_k (\boldsymbol{p})[n] \label{c0a}}
	{\label{OFa}}{\textbf{P1:}}
    \addConstraint{\sum_{k=1}^{K} ||\mathbf{A}[n] \mathbf{G}_{M,B}[n]||^2 p_k + \dot{\sigma}^2 ||\mathbf{A}[n]||_F^2 \leq P_{\mathrm{RIS}}[n]}\label{c1a}
    \addConstraint{0 \leq p_k[n] \leq p_k^{\max}[n], \forall{k} \in \mathcal{K}, \forall{n} \in \mathcal{N}}\label{c2a}
    \addConstraint{t_k^{\mathrm{loc}}[n] + t_k^{\mathrm{off}}[n] \leq D_k^{\max}, \forall{k} \in \mathcal{K}, \forall{m} \in \mathcal{M}}\label{c3a}
\end{mini!}

Problem \textbf{P1} remains non-convex due to the presence of the multi-user interference. Sequential fractional programming is an effective method for solving problems because it can handle interference without needing to eliminate it. This allows for complete resource utilization and avoids the negative consequences of noise enhancement \cite{zappone2017energy}. However, the key problem is to discover appropriate approximations of the denominator of objective function (\ref{c0a}) for any value of $k$. To begin with, we need to consider to find the upper-bound of the logarithm function \cite{papandriopoulos2006low} as follows.
\begin{equation}\label{upperbound}
    \log_2(1+\gamma) \leq \widetilde{a} \log_2 \gamma + \widetilde{b},
\end{equation}
where
\begin{equation}\label{abupdate}
    \widetilde{a} = \frac{\widetilde{\gamma}}{1+\widetilde{\gamma}}, \widetilde{b} = \log_2(1+\widetilde{\gamma}) - \frac{\widetilde{\gamma}}{1+\widetilde{\gamma}} \log_2 \widetilde{\gamma},
\end{equation}
with $\widetilde{\gamma}$ being any positive real number. At the point where $\gamma = \widetilde{\gamma}$, both the right-hand side (RHS) and left-hand side (LHS) of (\ref{upperbound}) are identical, and this equality extends to their derivatives with respect to $\gamma$ when evaluated at $\gamma = \widetilde{\gamma}$. Hereby, by applying (\ref{upperbound}), the upper-bound of objective function (\ref{c0a}) can be derived as in (\ref{derupper}). Similarly, this method can be applied to constraint (\ref{c3a}) to find its convex upper-bound.
\begin{figure*}[t!]
\begin{align}\label{derupper}
&\begin{aligned}
    &E^{\mathrm{Tot}}_k (\boldsymbol{p}) [n]=  E^{\mathrm{loc}}_k[n] +  E^{\mathrm{off}}_k (\boldsymbol{p}) [n] =  E^{\mathrm{loc}}_k[n] + \frac{p_k[n] o_k[n] S_k[n]}{W \log_{2}(1+\gamma_k[n])} \leq  E^{\mathrm{loc}}_k[n] + \frac{ p_k[n] o_k[n] S_k[n]}{ W [\widetilde{a}_k \log_2 \gamma_k[n] + \widetilde{b}_k]}=\\
    & E^{\mathrm{loc}}_k[n] + \frac{p_k[n] o_k[n] S_k[n]}{W [\widetilde{b}_k + \widetilde{a}_k \log_2 (|\mathbf{g}_k[n]|^2 p_k[n])- \widetilde{a}_k \log_2(\sum_{i \neq k}^K |\mathbf{g}_j[n]|^2 p_j[n] + \dot{\sigma}^2|| \mathbf{\Theta}_\lambda[n] \mathbf{A}[n] \mathbf{G}_{M,B}[n] ||^2 + \ddot{\sigma}^2)]} = \widetilde{E}^{\mathrm{Tot}}_k (\boldsymbol{p}) [n].
\end{aligned}
\end{align}
\end{figure*}

Afterwards, by replacing $p_k = 2^{q_k}$, we obtain as in (\ref{substitute}).
\begin{figure*}
\begin{equation}\label{substitute}
    \widetilde{E}^{\mathrm{Tot}}_k (\boldsymbol{q})[n]= E^{\mathrm{loc}}_k[n] + \frac{ 2^{q_k}[n] o_k[n] S_k[n]}{W [\widetilde{b}_k + \widetilde{a}_k \log_2 (|\mathbf{g}_k[n]|^2) + \widetilde{a}_k q_k[n]- \widetilde{a}_k \log_2(\sum_{i \neq k}^K |\mathbf{g}_j[n]|^2 2^{q_j}[n] + \dot{\sigma}^2|| \mathbf{\Theta}_\lambda[n] \mathbf{A}[n] \mathbf{G}_{M,B}[n] ||^2 + \ddot{\sigma}^2)]}.
\end{equation}
\hrulefill
\end{figure*}
Now, we can substitute (\ref{substitute}) as the upper bound function for our objective function (\ref{c0a}), allowing us to replace problem \textbf{P1} as follows.
\begin{mini!}|b|[1]
	{\boldsymbol{q}} {\sum_{n=1}^{N} \sum_{k=1}^{K} \widetilde{E}^{\mathrm{Tot}}_k (\boldsymbol{q})[n] \label{c0a1}}
	{\label{OFa1}}{\textbf{P11:}}
    \addConstraint{\sum_{k=1}^{K} ||\mathbf{A}[n] \mathbf{G}_{M,B}[n]||^2 2^{q_k} + \dot{\sigma}^2 ||\mathbf{A}[n]||_F^2 \leq P_{\mathrm{RIS}}[n]}\label{c1a1}
    \addConstraint{0 \leq 2^{q_k}[n] \leq p_k^{\max}[n], \forall{k} \in \mathcal{K}, \forall{n} \in \mathcal{N}}\label{c2a1}
    \addConstraint{t_k^{\mathrm{loc}}[n] + \frac{o_k[n] S_k[n]}{f_d(\boldsymbol{q})} \leq D_k^{\max}, \forall{k} \in \mathcal{K}, \forall{m} \in \mathcal{M},}\label{c3a1}
\end{mini!}
where $f_d(\boldsymbol{q}) = W[\widetilde{b}_k + \widetilde{a}_k \log_2 (|\mathbf{g}_k[n]|^2) + \widetilde{a}_k q_k[n] - \widetilde{a}_k \log_2(\sum_{i \neq k}^K |\mathbf{g}_j[n]|^2 2^{q_j}[n] + \dot{\sigma}^2|| \mathbf{\Theta}_\lambda[n] \mathbf{A}[n] \mathbf{G}_{M,B}[n] ||^2 + \ddot{\sigma}^2)]$. Since approximate problem $\textbf{P11}$ represents a fractional problem, we can utilize quadratic transformation techniques to effectively resolve this issue \cite{shen2018fractional}. Moreover, (\ref{c0a1}) has convex numerator and concave denominator for any given $\tilde{a}_k$ and $\tilde{b}_k$. By introducing the auxiliary variable $\boldsymbol{\eta} = [\eta_1, \eta_2, \dots, \eta_k]$ to (\ref{c0a1}), we have the new objective function as
\begin{align}
&\begin{aligned}
    &\widetilde{E}^{\mathrm{Tot}}_k (\boldsymbol{q}, \boldsymbol{\eta})[n] = E^{\mathrm{loc}}_k[n] + 2 \eta_k \sqrt{2^{q_k}[n] o_k[n] S_k[n]}- \\
    &{\eta_k}^2 \left(W [\widetilde{b}_k + \widetilde{a}_k \log_2 (|\mathbf{g}_k[n]|^2) + \widetilde{a}_k q_k[n]- \right. \\
    &\left. {\resizebox{.94\hsize}{!}{$ \widetilde{a}_k \log_2(\sum_{i \neq k}^K |\mathbf{g}_j[n]|^2 2^{q_j}[n] + \dot{\sigma}^2|| \mathbf{\Theta}_\lambda[n] \mathbf{A}[n] \mathbf{G}_{M,B}[n] ||^2 + \ddot{\sigma}^2)] $}} \vphantom{W [\widetilde{b}_k} \right).\\
\end{aligned}
\end{align}

Hereby, we can reformulate problem $\textbf{P11:}$ as
\begin{mini!}|b|[2]
	{\boldsymbol{q}, \boldsymbol{\eta}} {\sum_{n=1}^{N} \sum_{k=1}^{K} \widetilde{E}^{\mathrm{Tot}}_k (\boldsymbol{q}, \boldsymbol{\eta})[n] \label{c0a1a}}
	{\label{OFa1a}}{\textbf{P12:}}
    \addConstraint{\text{(\ref{c1a1})}, \text{(\ref{c2a1})}}\label{c1a1a}
    \addConstraint{\boldsymbol{\eta} \in \mathbb{R}, \forall{k} \in \mathcal{K}.}\label{c2a1a}
\end{mini!}

We proceed to optimize the variables $q_k$ and $\eta_k$ in an iterative manner. When $q_k$ is fixed, the optimal $\eta_k^*$ can be obtained in closed form as
\begin{equation}\label{etaupdate}
    \eta_k^* = \frac{\sqrt{2^{q_k}[n] o_k[n] S_k[n]}}{f_d(\boldsymbol{q})}.
\end{equation}
Consequently, for fixed $\eta_k$, getting the optimal $q_k^*$ is a convex optimization problem since log-sum-exp function is convex \cite{boyd2004convex}, which we can solve by employing CVXPY solver library in Python programming. Hence, sequential fractional programming algorithm for transmit power control problem is established in Algorithm \ref{algo1}.
\begin{algorithm}[t]
\caption{Sequential Fractional Programming for Transmit Power Control}
\label{algo1}
\SetKwInOut{Input}{Input}
\SetKwInOut{Output}{Output}
\Input{Iteration index $\tau = 0$, $\tau_{\max}$, initial any feasible $\boldsymbol{p}^0$, tolerance value $\varepsilon > 0$}
\BlankLine
\textbf{Initialization:} Set $\widetilde{\gamma}_k^0 = \gamma_k^0(\boldsymbol{p}^0)$ and compute $\widetilde{a}_k^0$ and $\widetilde{b}_k^0$ by (\ref{abupdate})\;
\Repeat{}{
    Set $\tau = \tau + 1$\;
    Update $\boldsymbol{\eta}^{\tau}$ by (\ref{etaupdate})\;
    Update $\boldsymbol{q}^{\tau}$ with parameters $\widetilde{a}_k^{(\tau-1)}$ and $\widetilde{b}_k^{(\tau-1)}$ by solving the convex optimization\;
    Compute $\boldsymbol{p} = 2^{\boldsymbol{q}}$\;
    Compute $\widetilde{\gamma}_k^{\tau} = \gamma_k^{\tau}(\boldsymbol{p})$\;
    Update $\widetilde{a}_k^{\tau}$ and $\widetilde{b}_k^{\tau}$ by (\ref{abupdate})\;
}{
$|E^{\mathrm{Tot}}_k(\tau) - E^{\mathrm{Tot}}_k(\tau-1)| \leq \varepsilon$ or $\tau > \tau_{\max}$\;
}
\Output{Optimal Transmit Power Control $\boldsymbol{p}^*$}
\end{algorithm}

\subsection{Partial Task Offloading Problem}\label{ptop}
With fixed $\boldsymbol{\beta}_{\lambda}, \boldsymbol{\phi}_{\lambda}, \boldsymbol{\alpha}, \boldsymbol{j}, \boldsymbol{p}$, the second sub-problem can be given as
\begin{mini!}|b|[1]
	{\boldsymbol{o}} {\sum_{n=1}^{N} \sum_{k=1}^{K}E^{\mathrm{Tot}}_k (\boldsymbol{o})[n] \label{c0b}}
	{\label{OFb}}{\textbf{P2:}}
    \addConstraint{t_k^{\mathrm{loc}}[n] + t_k^{\mathrm{off}}[n] \leq D_k^{\max}, \forall{k} \in \mathcal{K}, \forall{m} \in \mathcal{M}}\label{c1b}
    \addConstraint{\sum_{k=1}^{K} o_k[n] S_k[n] C_k \leq C^0, \forall{m} \in \mathcal{M}}\label{c2b}
    \addConstraint{0 \leq o_k[n] \leq 1, \forall{k} \in \mathcal{K}, \forall{n} \in \mathcal{N}.}\label{c3b} 
\end{mini!}

Since the objective function in problem $\textbf{P2}$ is linear and every constraint is either linear or affine, the problem is naturally convex. Therefore, it is suitable to apply convex optimization techniques to find the optimal $\boldsymbol{o}^*$, and we can employ CVXPY solver library in Python programming.

\subsection{Joint Amplitude, Phase Shift, Amplification, and Task Admission Problem}\label{jointprob}
When data arrivals are stochastic, it becomes difficult to achieve the long-term queue stability constraint (\ref{c7}) when the decisions are made in each time slot without further knowledge. Therefore, we employed the Lyapunov optimization framework to address the constraint (\ref{c7}) and estabilish a connection between the energy minimization problem into the queue stability \cite{li2019dynamic}.

First of all, a quadratic function of queue backlog size is used as a Lyapunov candidate function \cite{zhuang2020adaptive}, which is as follows.
\begin{equation}
    L(Q[n]) = \frac{1}{2} \sum_{k=1}^{K} Q_k[n]^2.
\end{equation}
With the definition of the Lyapunov function, the Lyapunov drift can be derived as
\begin{align}
&\begin{aligned}
    &\Delta(Q[n]) =L(Q[n+1])-L(Q[n]) \\
    &=\frac{1}{2}\sum_{k=1}^{K}[Q_k^2[n+1] - Q_k^2[n]]\\
    &=\frac{1}{2}\sum_{k=1}^{K}[(\max\{Q_k[n]-S_k[n] ,0\}+j_k[n])^2-Q_k^2[n]]\\
    &\leq\frac{1}{2}\sum_{k=1}^{K}\frac{j_k^2[n]+S_k^2[n]}{2}+\sum_{k=1}^{K}Q_k[n][j_k[n]-S_k[n]]\\
\end{aligned}
\end{align}
where $(\max\{Q_k[n]-S_k[n] ,0\}+j_k[n])^2 \leq Q_k^2[n]+j_k^2[n]+S_k^2[n]+2Q_k[n][j_k[n]-S_k[n]]$. Through minimizing the Lyapunov drift at each time slot t, the system can be stabilized, resulting in the reduction of queue backlog and alleviating congestion. By combining with our objective function, we can define the drift-plus-penalty function and transform the corresponding optimization problem as
\begin{equation}
    \resizebox{\hsize}{!}{$
    \ddot{E}(\boldsymbol{\beta}_{\lambda}, \boldsymbol{\phi}_{\lambda}, \boldsymbol{\alpha}, \boldsymbol{j})[n] = \Delta(Q[n])+V\mathbb{E}(E^{\mathrm{Tot}}_k(\boldsymbol{\beta}_{\lambda}, \boldsymbol{\phi}_{\lambda}, \boldsymbol{\alpha}, \boldsymbol{j}) [n]|Q(t)),
    $}
\end{equation}
where $V$ is the non-negative control parameter to adjust between system stability and penalty reduction.

Under Lyapunov optimization framework with fixed $\boldsymbol{o}$ and $\boldsymbol{p}$, the third sub-problem can be expressed as
\begin{mini!}|b|[1]
	{\boldsymbol{\beta}_{\lambda}, \boldsymbol{\phi}_{\lambda}, \boldsymbol{\alpha}, \boldsymbol{j}}  {\sum_{n=1}^{N} \sum_{k=1}^{K} \ddot{E}(\boldsymbol{\beta}_{\lambda}, \boldsymbol{\phi}_{\lambda}, \boldsymbol{\alpha}, \boldsymbol{j})[n] \label{c0c}}
	{\label{OFc}}{\textbf{P3:}}
    \addConstraint{\sum_{k=1}^{K} ||\mathbf{A}[n] \mathbf{G}_{M,B}[n]||^2 p_k + \dot{\sigma}^2 ||\mathbf{A}[n]||_F^2 \leq P_{\mathrm{RIS}}[n]}\label{c1c}
	\addConstraint{t_k^{\mathrm{loc}}[n] + t_k^{\mathrm{off}}[n] \leq D_k^{\max}, \forall{k} \in \mathcal{K}, \forall{m} \in \mathcal{M}}\label{c2c}
    \addConstraint{j_k[n] \leq J_k[n], \forall{k} \in \mathcal{K}, \forall{n} \in \mathcal{N}}\label{c3c}
    \addConstraint{S_k[n] \leq Q_k[n], \forall{k} \in \mathcal{K}, \forall{n} \in \mathcal{N}}\label{c4c}
    \addConstraint{\beta^{r}_m[n] + \beta^{t}_m[n] = 1, \forall{m} \in \mathcal{M}, \forall{n} \in \mathcal{N}}\label{c6c}
    \addConstraint{0 \leq \phi^{t}_m[n], \phi^{r}_m[n] < 2\pi, \forall{m} \in \mathcal{M}, \forall{n} \in \mathcal{N}.}\label{c7c}
\end{mini!}

Due to the non-convex nature in both the complex objective function and non-linear constraint (\ref{c1c}) of problem $\textbf{P3}$, traditional convex optimization cannot handle effectively. DRL is a suitable solution approach due to its ability to handle high-dimensional, non-linear, and complex decision spaces effectively \cite{luong2019applications}. In contrast to traditional optimization techniques, DRL does not require the problem to be convex and can learn optimal policies directly from interactions with the environment. This makes it well-suited for problems where explicit modeling is challenging or infeasible.

Among DRL methods, DDQN is particularly suitable for several reasons. DDQN effectively addresses the overestimation bias present in standard $\mathcal{Q}$-learning by decoupling action selection and value estimation, resulting in more stable and accurate learning \cite{van2016deep}. This is crucial for environments with a large number of possible actions and uncertain outcomes, as it ensures more reliable value approximations.

Initially, DRL is conceptualized as a Markov decision process (MDP), which is described by a 4-tuple including state $s$, action $a$, immediate rewards $\mathcal{R}$, and a transition function $\mathcal{P}$. During each time step, the agent applies a policy function $\hat{\pi}(a|s)$ to choose an action depending on its present state. The agent then receives a reward from the environment. This stochastic policy, $\hat{\pi}(a|s) = \mathrm{Pr}(A_n=a|S_n=s)$, represents the likelihood of selecting action $a$ in state $s$ at time $n$. The objective of the agent is to develop a policy that optimizes the expected total reward over time, which can be expressed mathematically as the sum of cumulative rewards, $\max_{\hat{\pi}} \mathbb{E} \sum_{i=n}^{\infty}(\xi^{i-n} \mathcal{R}_i)$, with $0<\xi<1$ being the discount factor that determines the extent to which future benefits are valued compared to immediate rewards, hence influencing the agents' preference between short-term and long-term gains.


In order to implement DDQN, we model our active STAR-RIS-assisted MEC system as MDP.
\subsubsection{Agent}The BS functions as an agent in our system, continuously interacting with an environment by taking actions based on its current state and receiving feedback through rewards.

\subsubsection{Environment} The agent interacts with the system, which is known as its environment. Its components include the BS, active STAR-RIS, MEC server, and channel models. The environment transitions to new states based on the actions taken, altering the conditions and rewards the agent encounters, impacting its future decisions and learning process.

\subsubsection{State} The state encompasses all relevant information about the environment that the agent can perceive at a specific time step. The state continually updates in response to the agent's actions and inherent environmental changes. In our system, the state $\boldsymbol{s}_n$ at time $n$ includes $\{\mathbf{g}_k[n], \boldsymbol{\beta}_{\lambda}[n], \boldsymbol{\phi}_{\lambda}[n], \boldsymbol{\alpha}[n], Q_k[n], k \in \mathcal{K}, \lambda \in \{r, t\}\}$.

\subsubsection{Action} The action is a decision made by the agent that influences the state of the environment. The chosen action leads to a transition in the environment, resulting in new states and rewards that inform the future decisions for the agent. In our system, the action $\boldsymbol{a}_n$ at time $n$ includes amplitude vector $\boldsymbol{\beta}_{\lambda}$, phase shift coefficient vector $\boldsymbol{\phi}_{\lambda}$, amplification coefficient vector $\boldsymbol{\alpha}$, admitted task vector $\boldsymbol{j}$, and are defined as the incremental values of the current values as follows.
\begin{equation}
    \boldsymbol{\beta}_{\lambda}[n+1] = \boldsymbol{\beta}_{\lambda}[n+1] \odot \triangle \boldsymbol{\beta}_{\lambda}[n],
\end{equation}
\begin{equation}
    \boldsymbol{\phi}_{\lambda}[n+1] = \boldsymbol{\phi}_{\lambda}[n+1] \odot \triangle \boldsymbol{\phi}_{\lambda}[n],
\end{equation}
\begin{equation}
    \boldsymbol{\alpha}[n+1] = \boldsymbol{\alpha}[n+1] \odot \triangle \boldsymbol{\alpha}[n],
\end{equation}
\begin{equation}
    \boldsymbol{j}[n+1] = \boldsymbol{j}[n+1] \odot \triangle \boldsymbol{j}[n],
\end{equation}
where $\odot$ represents the Hadamard (element-wise) product, and $\triangle$ represents the incremental value at time $n$. 

\subsubsection{Transition Function} The transition function, $\mathcal{P}(s'|s,a)$ is the probability of moving to a new state $s'$ given the current state $s$ and action $a$.

\subsubsection{Reward} The reward function at time slot $n$, $\mathcal{R}_n(s,a,s')$ reflects the immediate benefit or cost of performing a specific action $\boldsymbol{a}_n=a$ in state $\boldsymbol{s}_n=s$ and transitioning to state $\boldsymbol{s}_{n+1}=s'$, which guides the agent's learning process by providing feedback on the desirability of actions. Our system model aims to minimize the drift-plus-penalty function while simultaneously satisfying the QoS constraints, including power budget and tolerable latency for task completion. Accordingly, we establish our reward function as follows.
\begin{align}
&\begin{aligned}
    & \mathcal{R}_n(s,a,s') = \sum_{k=1}^{K} z_1 \ddot{E} [n] + \sum_{k=1}^{K} z_2 F(Q_k[n] - S_k[n]) + \\
    & z_3 F(P_{\mathrm{RIS}}[n] -  \sum_{k=1}^{K} (||\mathbf{A}[n] \mathbf{G}_{M,B}[n]||^2  p_k + \dot{\sigma}^2 ||\mathbf{A}[n]||_F^2))+ \\
    & \sum_{k=1}^{K} z_4 F(D_k^{\max} - t_k^{\mathrm{loc}}[n] - t_k^{\mathrm{off}}[n]) + \sum_{k=1}^{K} z_5 F(J_k[n] - j_k[n]), \\
\end{aligned}
\end{align}
where $z_1, z_2, z_3, z_4$ and $z_5$ are the weight coefficients for balancing the terms, and $F(x)$ is a piece-wise function defined as
\begin{equation}
 F(x) = \left \{ \begin{array}{ll}{p_0} & {\text{when } x \geq 0}, \\ {x,} & {\text {otherwise,}}\end{array}\right.
\end{equation}
where $p_0$ is positive constant to represent revenue. Furthermore, under the given policy $\hat{\pi}$, the action value function ($\mathcal{Q}$-function) can be expressed as
\begin{equation}
    \mathcal{Q}^{\hat{\pi}}(s,a) = \mathbb{E}^{\hat{\pi}}\left[ \mathcal{R}_1 + \xi \mathcal{R}_2 + \dots | \boldsymbol{s}_n=s, \boldsymbol{a}_n=a  \right].
\end{equation}

The $\mathcal{Q}$-function measures the long-term value of actions, enabling the agent to select actions that maximize immediate rewards and achieve energy efficiency and QoS satisfaction in future time slots. Hence, the optimal $\mathcal{Q}$-function, which is mathematically expressed as $\mathcal{Q}^*(s,a) = \max_{\hat{\pi}} \mathcal{Q}^{\hat{\pi}}(s,a)$ can be defined in terms of Bellman equation as
\begin{equation}
    \mathcal{Q}^*(s,a) = \sum_{s'} \mathcal{P}(s'|s,a) \left[\mathcal{R}(s,a,s') + \xi \max_{a'} \mathcal{Q}^*(s',a') \right].
\end{equation}
The optimal $\mathcal{Q}$-function $\mathcal{Q}^*(s,a)$ can be determined by utilizing a $\mathcal{Q}$-table. This $\mathcal{Q}$-table is iteratively updated through the $\mathcal{Q}$-learning algorithm based on the agent's experiences. However, when the number of states and actions increases, maintaining and updating a $\mathcal{Q}$-table becomes increasingly complex and resource-intensive. To overcome this, deep $\mathcal{Q}$-networks (DQNs) are employed.

\subsubsection{DQN} The DQN algorithm incorporates a deep neural network ($\mathcal{Q}$-network) to estimate $\mathcal{Q}$-values. The $\mathcal{Q}$-network with weight parameters $\boldsymbol{\omega}$, denoted as $\mathcal{Q}(s,a;\boldsymbol{\omega})$, receives the state $s$ as input and generates the $\mathcal{Q}$-values as outputs for any available action $a$. The update rule for the $\mathcal{Q}$-value in DQN is given by
\begin{align}
&\begin{aligned}
    &\mathcal{Q}_{n+1}(s,a;\boldsymbol{\omega}) = \mathcal{Q}_n(s,a;\boldsymbol{\omega})+\\
    &\hat{l}\left[\mathcal{R}_n(s,a,s') + \xi \max_{a'} \mathcal{Q}_n^{\dag}(s',a';\boldsymbol{\omega}^{\dag})-\mathcal{Q}_n(s,a;\boldsymbol{\omega})\right],\\
\end{aligned}
\end{align}
where $0<\hat{l}\leq 1$ is the learning rate to control the step size during the weight update process, and $\mathcal{Q}^{\dag}$ denotes the target network with weight parameters $\boldsymbol{\omega}^{\dag}$ used to stabilize the updates. However, DQN is afflicted by overestimation bias due to the usage of same $\mathcal{Q}$-network for both action selection and evaluation. To overcome this issue, DDQN is implemented.

\subsubsection{DDQN} The DDQN algorithm decouples the process of selecting actions from evaluating actions by utilizing two distinct networks. The primary network is used to select the action that maximizes the $\mathcal{Q}$-value for next state, whereas target network is used to evaluate the $\mathcal{Q}$-value of the action selected by the primary network. This allows DDQN to mitigate the overestimation problem, leading to more accurate $\mathcal{Q}$-value estimates and improved policy performance. The update rule for the $\mathcal{Q}$-value in DDQN is given by
\begin{align}
&\begin{aligned}
    &\mathcal{Q}_{n+1}(s,a;\boldsymbol{\omega}) = \mathcal{Q}_n(s,a;\boldsymbol{\omega})+\hat{l}\left[\mathcal{R}_n(s,a,s') + \vphantom{W [\widetilde{b}_k}  \right.\\
    & \left. \xi \mathcal{Q}_n\left(s', \max_{a'} \mathcal{Q}_n^{\dag}(s',a;\boldsymbol{\omega});\boldsymbol{\omega}^{\dag}\right)-\mathcal{Q}_n(s,a;\boldsymbol{\omega})\right].\\
\end{aligned}
\end{align}
The objective of DDQN is to minimize the loss function by decreasing the difference between the predicted $\mathcal{Q}$-values and the target $\mathcal{Q}$-values.  This process facilitates the network in acquiring precise Q-values for state-action pairs, which are subsequently utilized to make optimal decisions within the environment. The loss function is defined in the following manner.
\begin{equation}
    \mathcal{L}(\boldsymbol{\omega}) = \mathbb{E}_{(s,a,r,s')}\left[(y^{\mathrm{DDQN}} - \mathcal{Q}(s,a;\boldsymbol{\omega}))^2\right],
\end{equation}
where
$$y^{\mathrm{DDQN}} = \mathcal{R}(s,a,s') + \xi \mathcal{Q}(s', \mathrm{argmax}_{a'} \mathcal{Q}^{\dag}(s',a;\boldsymbol{\omega});\boldsymbol{\omega}^{\dag}).$$

\begin{figure}[t]
	\includegraphics[width=\linewidth]{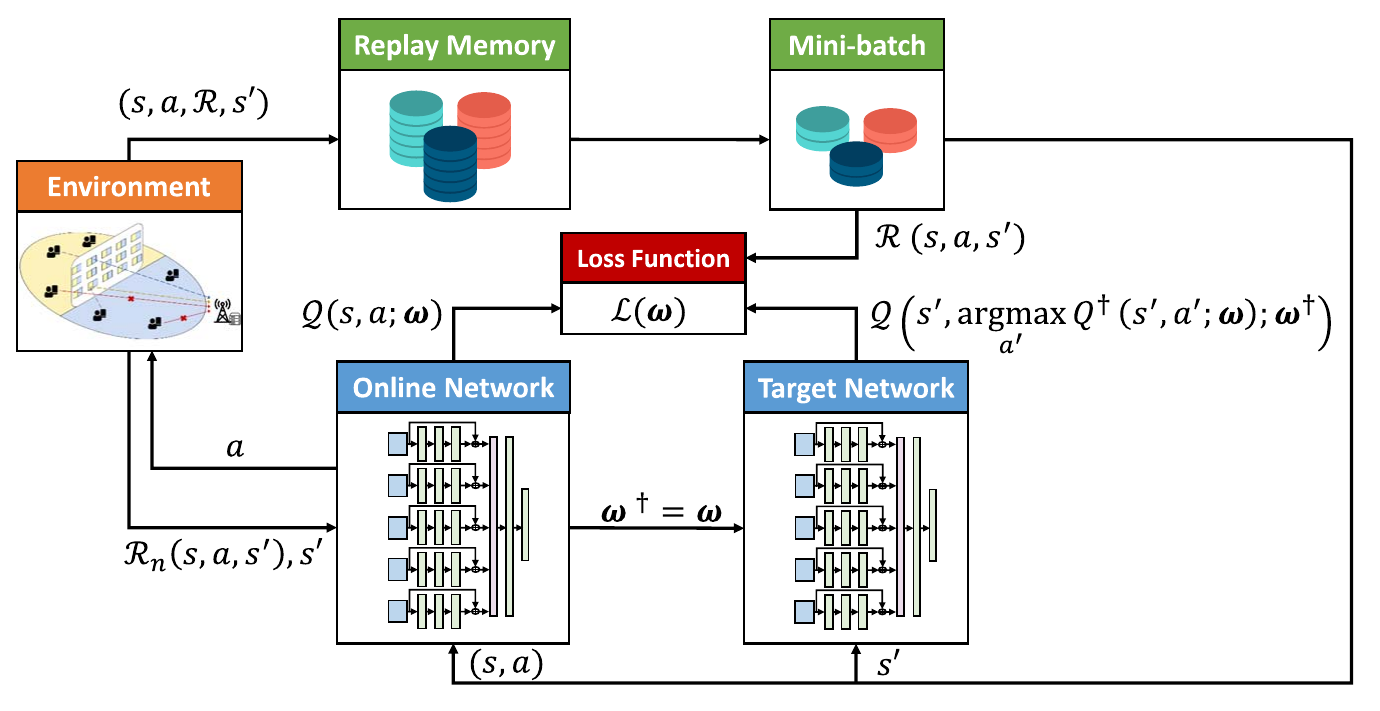}
	\caption{DDQN architecture.}
	\label{ddqn_archi}
\end{figure}

Fig. \ref{ddqn_archi} illustrates the DDQN approach for joint amplitude, phase shift, amplification, and task admission problem. For the online network, we utilize ResNet, incorporates a dense network to combine and process all input states, and employ the Swish activation function to enhance gradient flow and learning capacity. Consequently, the comprehensive description of the overall solution approach for our proposed active STAR-RIS-assisted MEC system is described in Algorithm \ref{algo2}.
\begin{algorithm}[t]
\caption{Combined Sequential Fractional, Convex Optimization and Lyapunov Optimization with DDQN Algorithm}
\label{algo2}
\SetKwInOut{Input}{Input}
\Input{Parameters $\xi$, $\hat{l}$, mini batch size, replay buffer}
\BlankLine
\textbf{Initialization:} Set random $\mathcal{Q}$-network weights $\boldsymbol{\omega}$, and $\mathcal{Q}^\dag$-network weights as $\boldsymbol{\omega}^{\dag}= \boldsymbol{\omega}$\;
\For{each episode $= 1, 2, \dots$}{
    Initialize the active STAR-RIS-assisted MEC environment and obtain initial observed states\;
    \For{each time slot $n=1, 2, \dots$}{
        Forward $\boldsymbol{p}[n]$ from \textbf{P1} and $\boldsymbol{o}[n]$ from \textbf{P2} along with the other observed states $\boldsymbol{s}_n$ into DDQN and obtain $\mathcal{Q}_n(s,a;\boldsymbol{\omega})$\;
        Select action $\boldsymbol{a}_n$, receive immediate reward $\mathcal{R}_n(s,a,s')$ and next state $\boldsymbol{s}_{n+1}$\;
        Store $(\boldsymbol{s}_n, \boldsymbol{a}_n, \mathcal{R}_n, \boldsymbol{s}_{n+1})$ into replay buffer\;
        Randomly select mini batch of $(\boldsymbol{s}_n, \boldsymbol{a}_n, \mathcal{R}_n, \boldsymbol{s}_{n+1})$ from replay buffer and calculate target values $\mathcal{R}_n(s,a,s')+ \xi \mathcal{Q}_n(s', \mathrm{argmax}_{a'} \mathcal{Q}_n^{\dag}(s',a;\boldsymbol{\omega});\boldsymbol{\omega}^{\dag})$\;
        Train DDQN and update weights $\boldsymbol{\omega}$\;
        Update target weights $\boldsymbol{\omega}^{\dag}=\boldsymbol{\omega}$ at every $n_{\mathcal{Q}}$ time slot\;
        Apply Algorithm \ref{algo1} to obtain $\boldsymbol{p}[n+1]$\;
        Apply CVXPY solver to obtain $\boldsymbol{o}[n+1]$\;
    }
}
\end{algorithm}

\subsection{Complexity Analysis}
The transmit power control problem is solved by implementing sequential fractional programming as outlined in Algorithm \ref{algo1}. The number of iterations required is $\mathcal{O}(K^2)$ due to the presence of $K$ variables. The computational complexity for the optimization method in the partial task offloading problem is $\mathcal{O}(K)$. For the DDQN method for joint amplitude, phase shift, amplification, and task admission problem, each forward pass through the neural network has a complexity dependent on the network architecture. For a network with $L$ layers, each with $n_l$ neurons, the computational complexity is represented as $\mathcal{O}(\sum_{l=1}^{L}n_{l-1} n_l)$. Hence, the overall computational complexity for Algorithm \ref{algo2} is portrayed as $\mathcal{O}(\ddot{T}(K^2+K + \sum_{l=1}^{L}n_{l-1} n_l)$, where $\ddot{T}$ denotes the number of iterations expected to achieve convergence.

\section{Performance Evaluation}\label{evaluation}
In this section, we evaluate the performance of our proposed active STAR-RIS-assisted MEC system via numerical analysis. The network system is modeled within a rectangular area region measuring 340 m $\times$ 100 m. It consists of the BS equipped with MEC server positioned at (0 m, 0 m, 10 m) in the Cartesian coordinate system. The STAR-RIS consisting of 32 elements is positioned at coordinates (-60 m, 50 m, 5 m) with uniform linear array of (0 m, 1 m, 0 m) along the Y-axis. There are a total of 20 users equipped with single antennas. While half of them are uniformly distributed in the region (-70 m, 0 m) $\times$ (-240 m, 100 m) which is referred to as the $T$ region, and half of them are uniformly distributed within (-50 m, 0 m) $\times$ (100 m, 100 m) which is referred to as the $R$ region. The simulation parameters are mentioned in Table \ref{simtab}.
\begin{table}
	\centering
	\caption{Simulation parameters}
	\begin{tabular}{|l|c|}
		\hline
		\textbf{Parameter}                 & \textbf{Value}    \\
		\hline \hline
		Maximum power budget for ARIS $P_{\mathrm{RIS}}$ & 10 dBm \cite{papazafeiropoulos2024two}\\ \hline
        Maximum power budget for each user $p_k^{\max}$ & 20 dBm \cite{papazafeiropoulos2024two} \\ \hline
        Rician factor $\hat{R}$      & 10 \cite{li2024improving} \\ \hline
		Bandwidth $W$		& 2 MHz \cite{li2024improving} \\ \hline
        Size of data input $S_k$		& [30, 150] Mbits \cite{aung2024aerial} \\ \hline
        Number of computation resources $C_k$		& 800 cycles/bits \cite{aung2024aerial} \\ \hline
        Maximum tolerable delay $D^{\max}_k$		& [1, 6] s \cite{aung2024aerial} \\ \hline
        Allocated computation frequency $f_k$		& [60, 180]  MHz \cite{aung2024aerial} \\ \hline
        Noise power $\sigma^2$		& -174 dBm \cite{li2024improving} \\ \hline
        Channel gain at reference distance $\mu$ & -40 dBm \\ \hline
        Path loss exponent $\delta$        & 4  \\ \hline
		Discount factor & 0.98 \\ \hline
        Initial exploration & 1 \\ \hline
        Exploration rate & 0.02 \\ \hline
		Learning rate & 0.001 \\ \hline
		Discount factor $\xi$ & 0.9 \\ \hline
		Mini batch size & 8 \\ \hline
		Number of episodes & 1,000 \\ \hline
        $z_1, z_2, z_3, z_4, z_5, V$ & 0.1, 0.8, 0.9, 0.9, 0.9, 100  \\ \hline
	\end{tabular}
	\label{simtab}
\end{table}
We compare the following benchmark schemes in order to validate our proposed system.
\subsubsection{Full Active STAR-RIS (Full A-STAR)} In this scheme, instead of partial offloading, all tasks are entirely offloaded to the MEC server through the active STAR-RIS (i.e., $o_k[n]=1$). The remaining control variables are solved by Algorithm \ref{algo2}.
\subsubsection{Passive STAR-RIS (P-STAR)} In this scheme, we consider the conventional nearly passive STAR-RIS without amplification coefficients, and the remaining control variables are solved by Algorithm \ref{algo2}.
\subsubsection{Active RIS (A-RIS)} In this scheme, we replace the active STAR-RIS with active RIS, which only supports the reflecting region (i.e., $\beta_m^r[n]=1$ and $\beta_m^t[n]=0$). The remaining control variables are solved by Algorithm \ref{algo2}.
\subsubsection{Passive RIS (P-RIS)} In this scheme, we consider the conventional nearly passive reflecting only RIS without amplification coefficients, and the remaining control variables are solved by Algorithm \ref{algo2}.
\begin{figure}[t]
	\includegraphics[width=\linewidth]{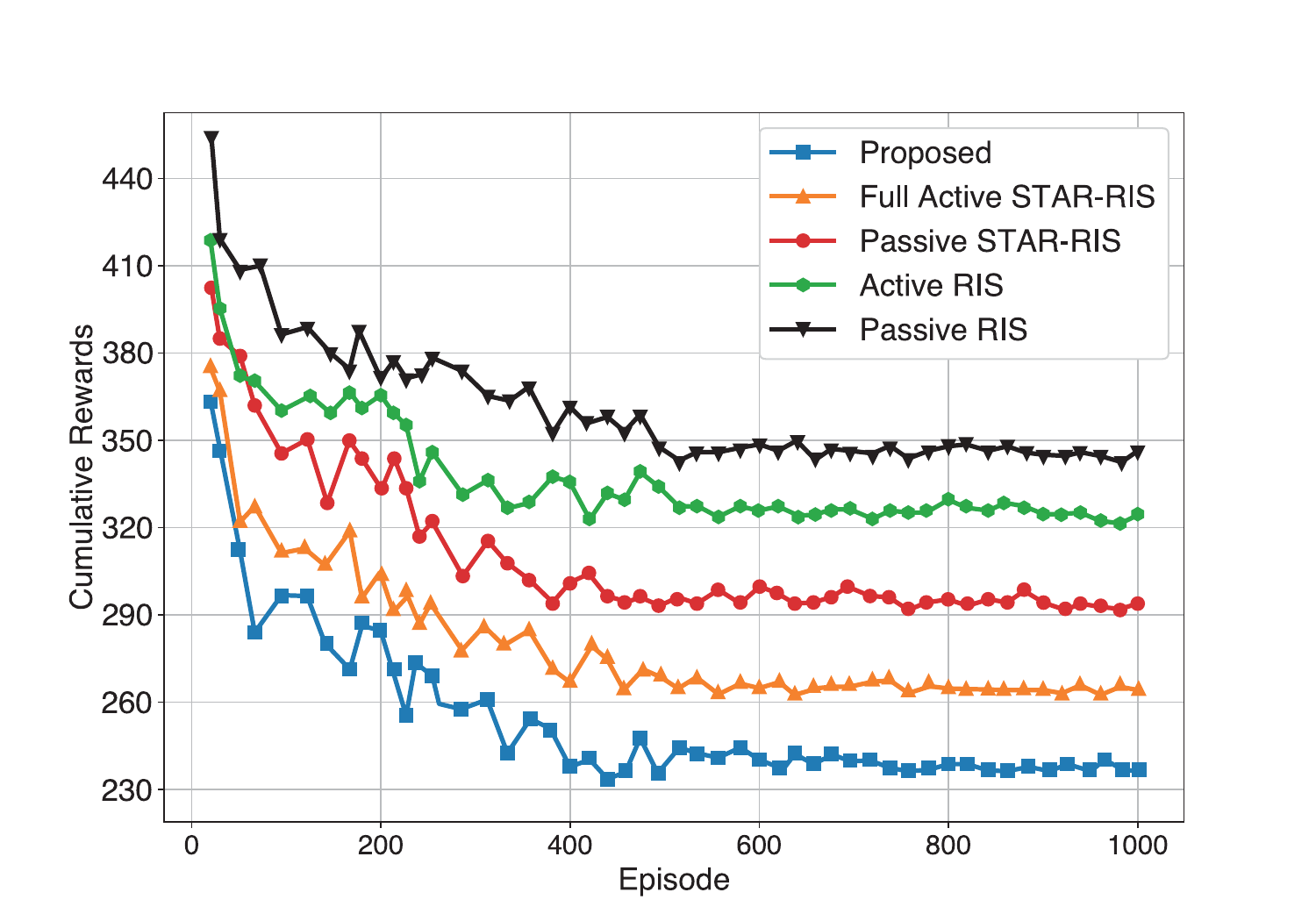}
	\caption{Performance comparison of cumulative rewards with different benchmark scenarios.}
	\label{reward}
\end{figure}

Fig. \ref{reward} shows the cumulative rewards obtained for our proposed scenario compared to the benchmark schemes. All the scenarios initially exhibit significant fluctuations but rapidly stabilize to achieve the cumulative rewards. The proposed scenario outperforms the benchmark schemes, achieving the lowest cumulative rewards and, therefore, the highest energy efficiency. This is because our proposed scheme employs a queue-management task-offloading mechanism and effectively utilizes the amplification and higher DoF of active STAR-RIS. This leads to the allocation of computational tasks between the user devices and the MEC server. Our proposed algorithm outperforms Full A-STAR by 9.43\%, P-STAR by 18.64\%, A-RIS by 26.15\%, and P-RIS by 30.43\%, respectively. 
\begin{figure}[t]
	\includegraphics[width=\linewidth]{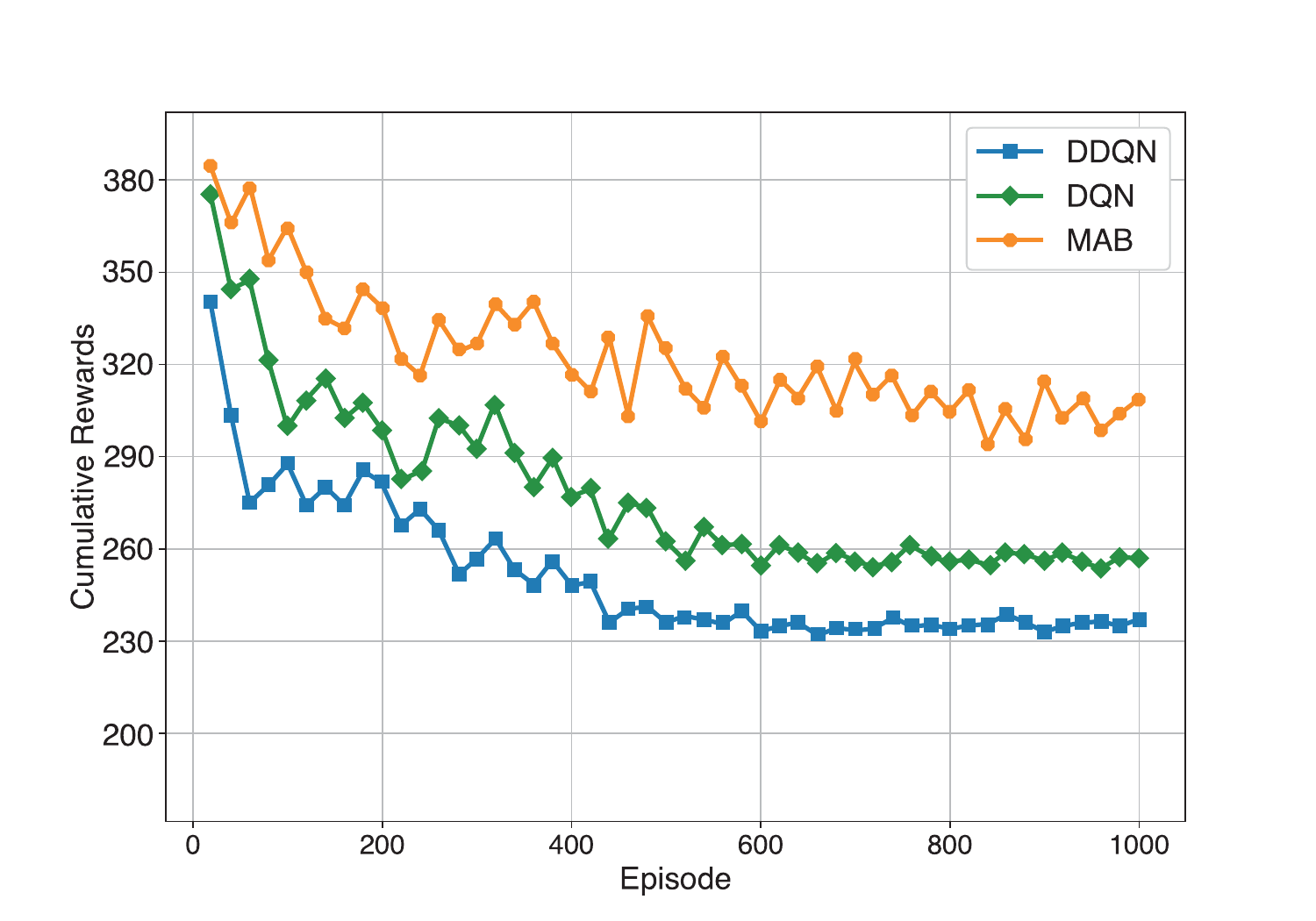}
	\caption{Performance comparison of cumulative rewards with different solution methods.}
	\label{methods}
\end{figure}

Fig. \ref{methods} compares the performance of DDQN to DQN and multiarmed bandit (MAB) algorithms. Here, MAB employs the epsilon-greedy method to maximize cumulative rewards by selecting a random arm with probability $\varsigma$ and the arm with the highest estimated reward with probability 1-$\varsigma$. Both DDQN and DQN methods surpass MAB because MAB prioritizes maximizing immediate rewards without considering future states, which is not ideal for complex environments that require a long-term strategy. Moreover, DDQN outperforms DQN by mitigating the overestimation bias through decoupling the action selection and evaluation processes. This leads to more precise Q-value estimates and improved overall performance. Here, DDQN algorithm outperforms the DQN algorithm by 5.88\%, and the MAB algorithm by 21.31\%.
\begin{figure}[t]
	\includegraphics[width=\linewidth]{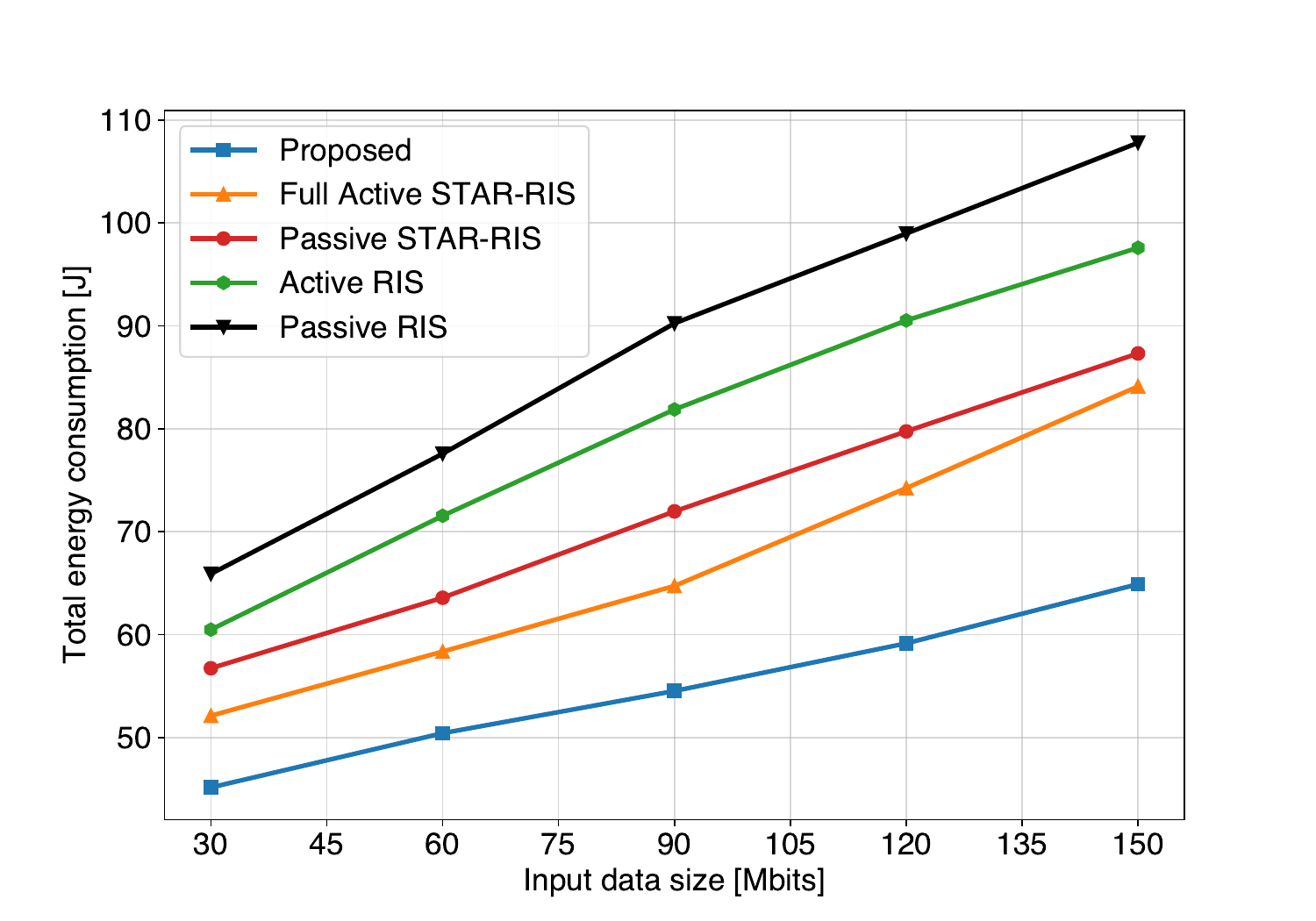}
	\caption{Performance comparison of total energy consumption under different input data size.}
	\label{ecvsinput}
\end{figure}

Fig. \ref{ecvsinput} illustrates the performance comparison of total energy consumption across different input data sizes, where $D^{\max}_k=$ 2 s. Our proposed active STAR-RIS approach consistently maintains the lowest total energy consumption as the input data size grows, highlighting its efficiency. This efficiency is attributed to the signal amplification capabilities that mitigate the multiplicative fading effect, thereby reducing the required transmission power for the users devices and lowering overall energy consumption. Additionally, the STAR-RIS schemes surpass the RIS schemes due to their ability to support users devices in both transmission and reflecting regions, leading to a higher DoF.
\begin{figure}[t]
	\includegraphics[width=\linewidth]{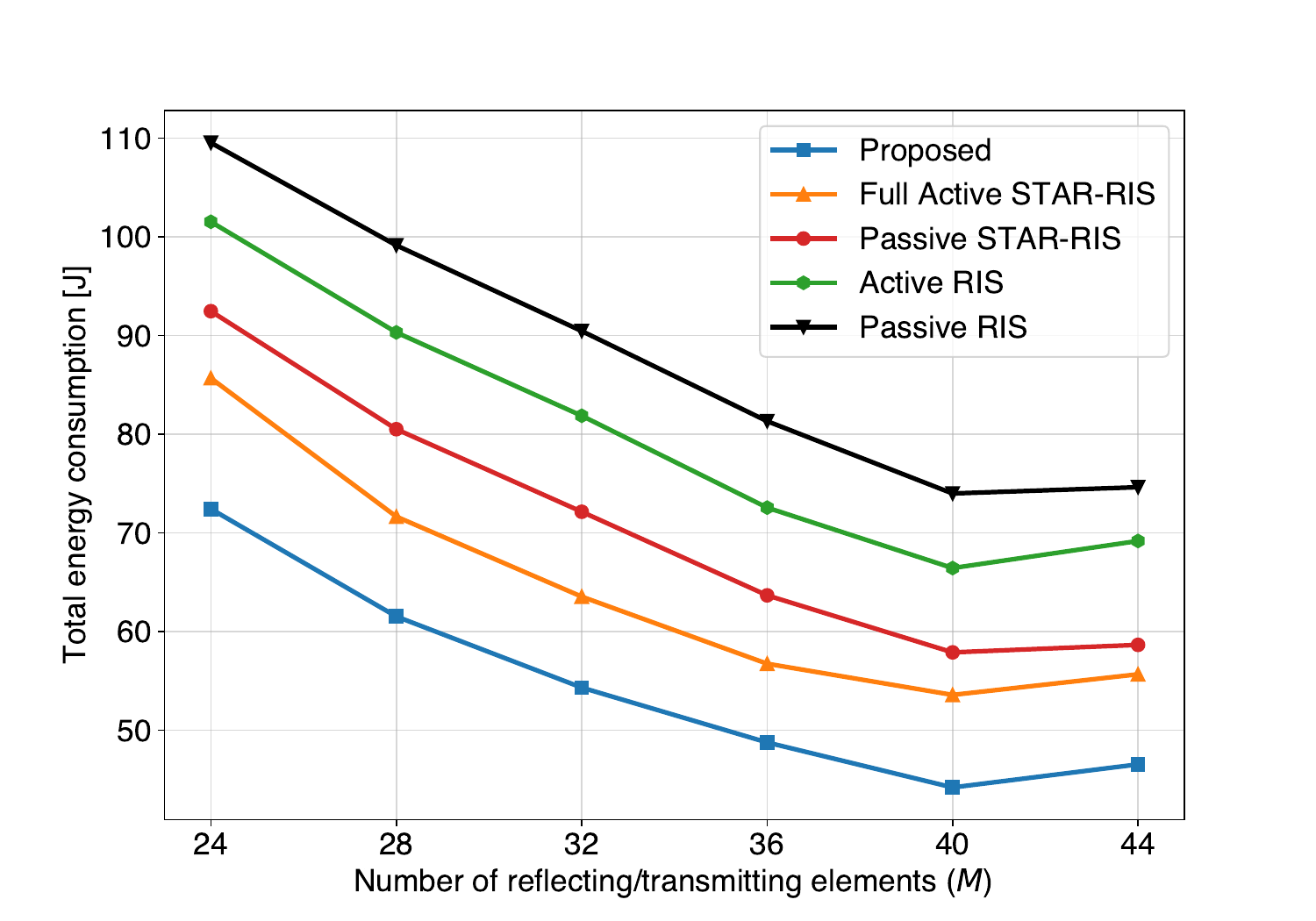}
	\caption{Performance comparison of total energy consumption under different number of reflecting/transmitting elements.}
	\label{ecvselements}
\end{figure}

Fig. \ref{ecvselements} demonstrates the performance comparison of total energy consumption across different numbers of reflecting/transmitting elements, where $S_k=$ 90 Mbits. Increasing the number of elements significantly impacts the overall energy consumption; however, the rate of reduction diminishes as the number of elements continues to grow. Therefore, it is unnecessary to deploy an excessively large number of components for a network comprising 20 user devices. The proposed active STAR-RIS system consistently outperforms benchmark schemes, exhibiting the lowest energy consumption. This performance advantage is primarily attributed to its enhanced signal amplification and task queue stability, which effectively mitigate the negative effects of multiplicative fading and optimize resource utilization.
\begin{figure}[t]
	\includegraphics[width=\linewidth]{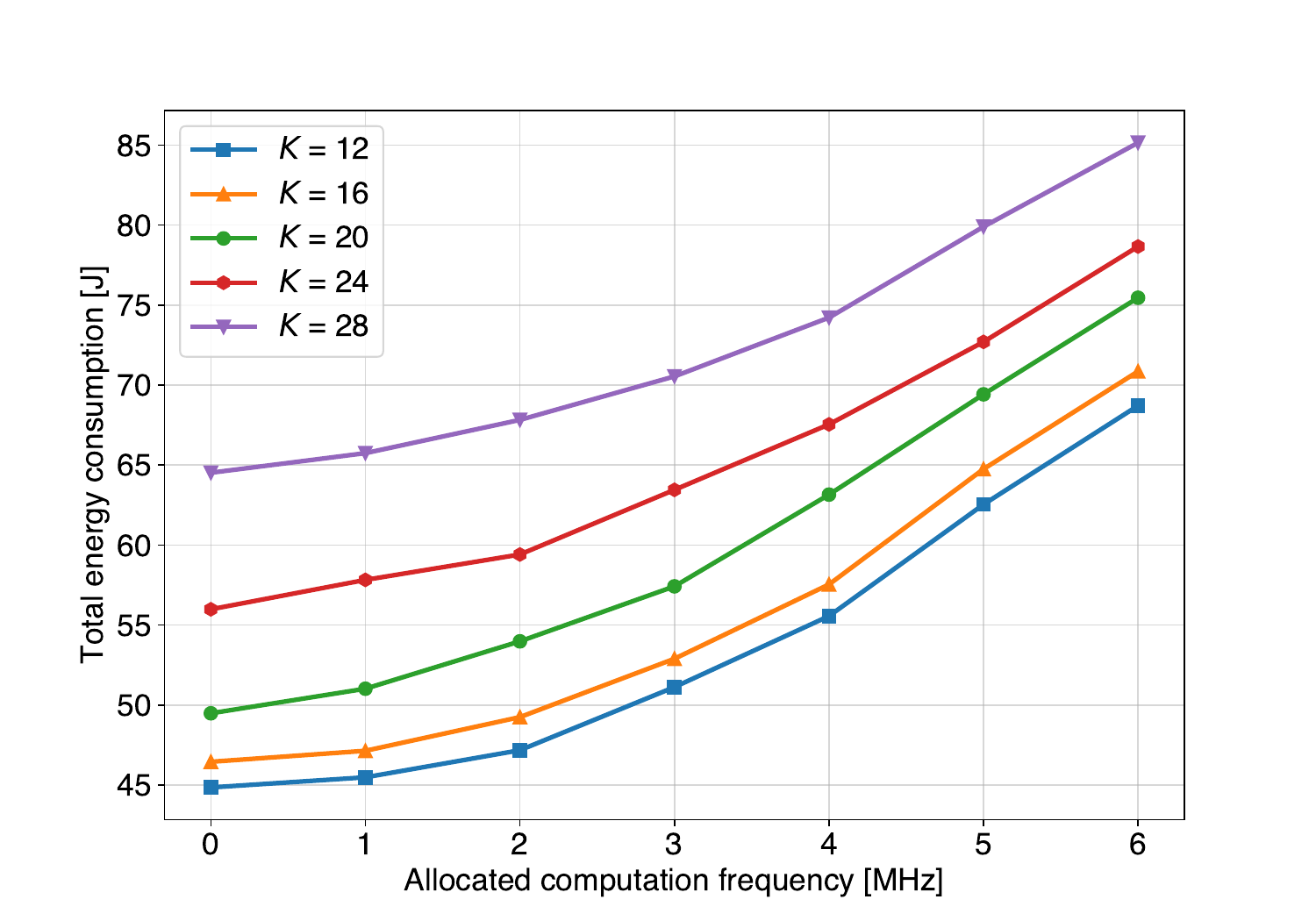}
	\caption{Performance comparison of total energy consumption under different allocated computation frequency.}
	\label{ecvscpu}
\end{figure}

Fig. \ref{ecvscpu} investigates the performance comparison of total energy consumption across allocated computation frequencies. The energy necessary to execute computations rises in association with the increase in the allocated computation frequency. Consequently, with the growing number of user devices, the total computational demand on the MEC system rises. Each additional user adds to the overall workload, thereby increasing the cumulative energy consumption. This highlights the importance of balancing computational performance with energy efficiency, particularly in scalable and sustainable MEC systems.
\begin{figure}[t]
	\includegraphics[width=\linewidth]{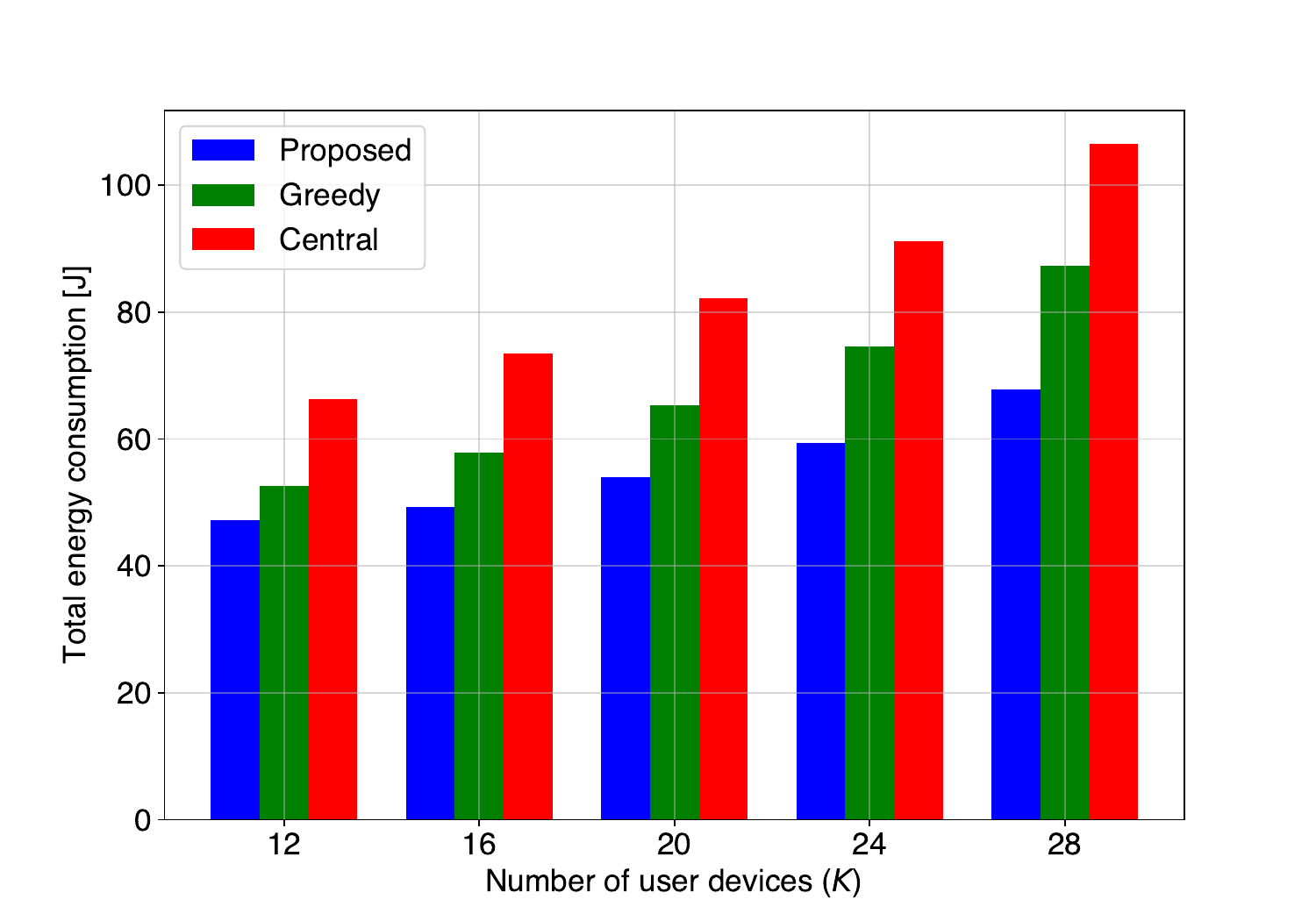}
	\caption{Performance comparison on impact of task queue management.}
	\label{queue}
\end{figure}

Fig. \ref{queue} examines the impact of task queue management under different numbers of user devices. We compared our proposed system with greedy offloading, where tasks are offloaded to the MEC server based on immediate gains without considering long-term queue stability, and a centralized scheme, where all tasks from different users are managed in a single, centralized queue at the MEC server. Our proposed system outperforms the benchmark schemes because it provides a more robust and efficient long-term optimization, adapting to changing network conditions and maintaining long-term queue stability. These advantages lead to better latency performance, which minimizes energy consumption.
\begin{figure}[t]
	\includegraphics[width=\linewidth]{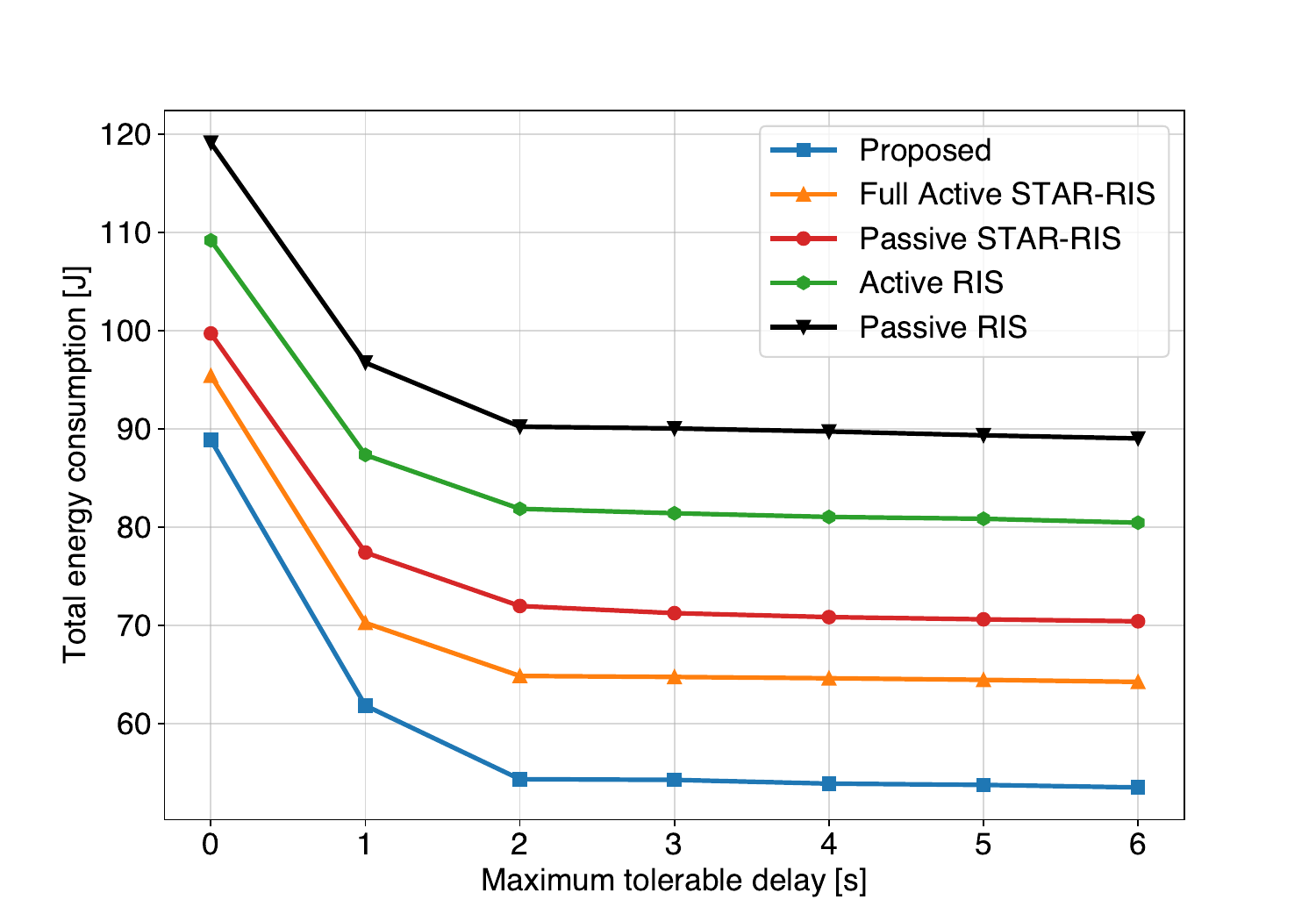}
	\caption{Performance comparison of total energy consumption under different maximum tolerable delay.}
	\label{ecvsdelay}
\end{figure}

Fig. \ref{ecvsdelay} represents the performance comparison of total energy consumption under different maximum tolerable delay. When the maximum tolerable delay is increased, the system has more time to complete the tasks for each user device. This results in the user devices operating at lower computation frequencies, efficient task scheduling, and reduction of the need for high-power transmission. Nevertheless, when the maximum tolerable delay continues to increase, the rate of decrease slows down, exhibiting characteristics of diminishing returns. This is because, beyond a certain threshold, the further increase in maximum tolerable delay provides less additional flexibility for energy savings. Among them, our proposed algorithm outperforms the benchmark schemes.
\begin{figure}[t]
	\includegraphics[width=\linewidth]{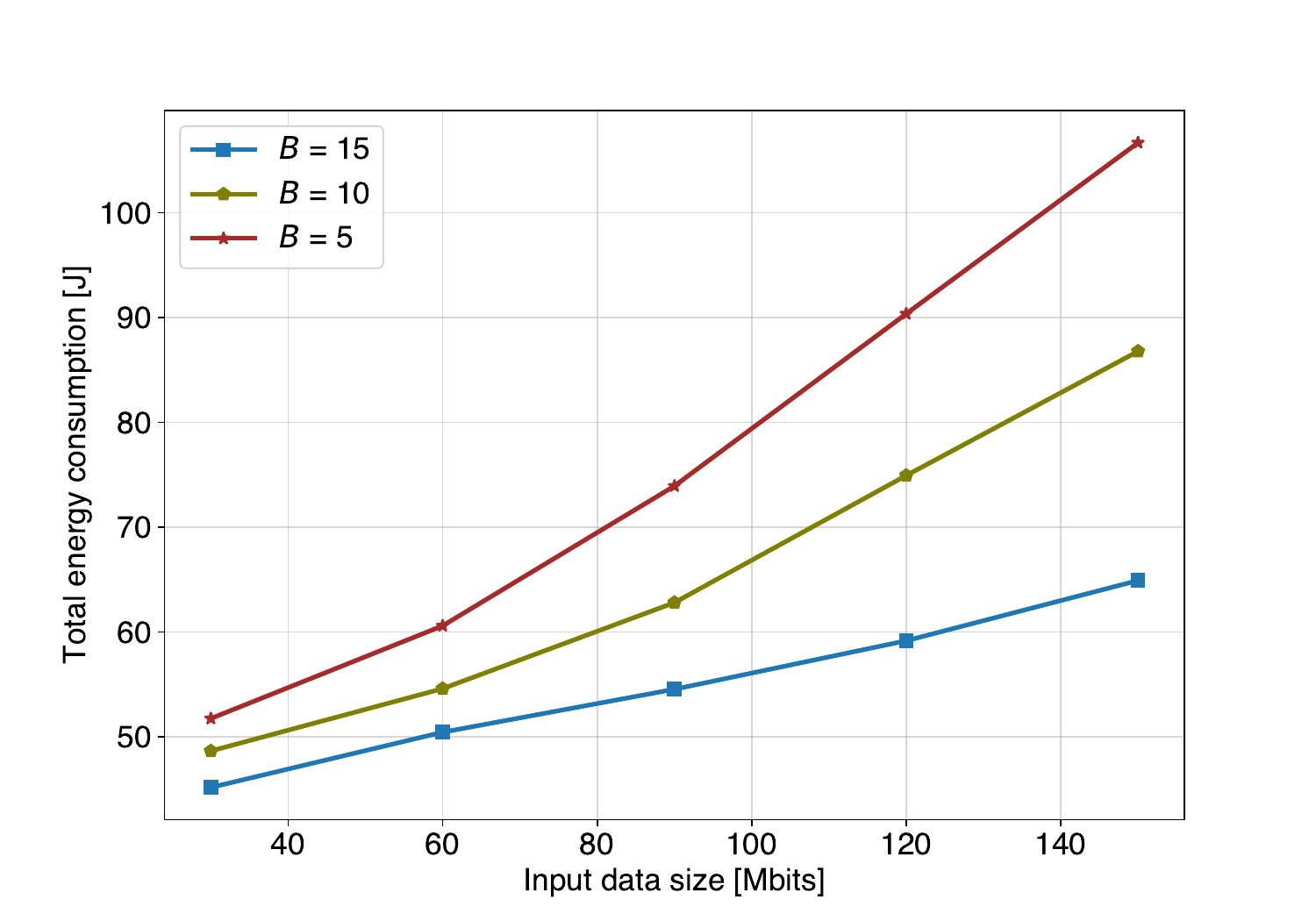}
	\caption{Performance comparison on impact of different number of antennas on total energy consumption.}
	\label{ecvsantenna}
\end{figure}

Fig. \ref{ecvsantenna} establishes the impact of different numbers of antennas. It is examined that increasing the number of antennas has a positive impact on reducing total energy consumption. This is primarily due to the benefits of improved signal quality, higher spectral efficiency, and enhanced beamforming capabilities. These factors result in an increase in channel gain for task offloading and mitigate the impact of inter-user interference. Consequently, systems equipped with a more significant number of antennas can transmit data more effectively and with less energy consumption, leading to improved overall energy efficiency.

\section{Conclusion}\label{conclusion}
In this article, we have studied an active STAR-RIS-assisted MEC system. To minimize the energy consumption of user devices, active STAR-RIS, and BS, we formulated a joint partial task offloading, amplitude, phase shift coefficients, amplification coefficients, transmit power of the BS, and admitted task problem. We decomposed the problem into three sub-problems due to its non-convexity, dynamic environment, and task arrivals. To solve these sub-problems, we implemented sequential fractional programming, convex optimization, and Lyapunov optimization with DDQN. Numerical findings indicate that our proposed system system surpasses the conventional STAR-RIS-assisted system by 18.64\% and the conventional RIS-assisted system by 30.43\%, respectively. This is due to the effective utilization of amplification and greater DoF of active STAR-RIS. Moreover, we conducted a comparison with the DQN and MAB methods to show the improvement due to the mitigation of the overestimation bias of the DDQN method. The performance is enhanced by 5.88\% and 21.31\% compared to the DQN and MAB methods, respectively. In addition, we compare with greedy offloading and centralized queue systems to demonstrate the benefits of incorporating task queue management and ensuring long-term queue stability. This approach enhances the latency performance and minimizes energy usage in the MEC system.
\bibliographystyle{IEEEtran}
\bibliography{mybib}

\end{document}